\RequirePackage{fix-cm}
\documentclass{iopjournal}

\RequirePackage{graphicx}
\RequirePackage{mathptmx}      
\RequirePackage{latexsym}
\usepackage{amsmath,amssymb,amsbsy,amstext,amsfonts}
\usepackage{bm}
\usepackage{subfigure}
\usepackage{tikz}
\usepackage{array}
\usepackage[title]{appendix}%

\usepackage[T1]{fontenc}
\usepackage{cite}

\def\D{\mathrm{d}}

\renewcommand{\vec}{\textbf}
\definecolor{LouisBlue}{RGB}{55, 114, 202}
\definecolor{LouisOrange}{RGB}{180, 54, 22}
\definecolor{LouisColor1}{RGB}{0, 118, 63}
\definecolor{LouisColor2}{RGB}{111, 73, 189}
\usetikzlibrary{matrix,shapes,fit,tikzmark,calc,3d,arrows.meta,scopes}
\usetikzlibrary{decorations.pathreplacing,decorations.markings,snakes,perspective,shadows,intersections}
\tikzset{middlearrow/.style={
		decoration={markings,
			mark= at position #1 with {\arrow{latex}} ,
		},
		postaction={decorate}
	}
}
\tikzset{inversemiddlearrow/.style={
		decoration={markings,
			mark= at position #1 with {-\arrow{latex[reversed]}} ,
		},
		postaction={decorate}
	}
}
\usepackage{hyperref}
\hypersetup{
	colorlinks=true,
	linkcolor=LouisBlue,
	citecolor=LouisBlue,
	urlcolor=LouisBlue,
} 

\begin{document}

\articletype{Paper} %	 e.g. Paper, Letter, Topical Review...

\title{Weaving the AdS spaces with partial-entanglement-entropy threads}

\author{Jiong Lin$^{1,*}$
        ,
        Yizhou Lu$^{2,3}$, 
         Qiang Wen$^{4}$,
         Yiwei Zhong$^{4,*}$}

\affil{$^1$School of Physics and Information Engineering, Guangdong University of Education, Guangzhou 510303, China}

\affil{$^2$Shanghai Institute for Mathematics and Interdisciplinary Sciences (SIMIS), Shanghai
200433, China}

\affil{$^3$Research Institute of Intelligent Complex Systems, Fudan University, Shanghai
200433, China}

\affil{$^4$Shing-Tung Yau Center and  School of Physics, Southeast University, Nanjing 210096, China}

\affil{$^*$These authors contribute equally.}

\email{luyz@simis.cn, wenqiang@seu.edu.cn}

\keywords{AdS-CFT Correspondence, Gauge-Gravity Correspondence, Models of
Quantum Gravity}

\begin{abstract}
In the context of the AdS/CFT correspondence, we propose a general scheme for reconstructing bulk geometric quantities in a static pure AdS background using the partial entanglement entropy (PEE), a measure of the entanglement structure on the boundary CFT.
The PEE between any two points $\mathcal{I}(\vec{x}, \vec{y})$ serves as the fundamental building block of the PEE structure. 
Any two-point PEE $\mathcal{I}(\vec{x}, \vec{y})$ can be geometrized by the bulk geodesic connecting two boundary points $\vec{x}$ and $\vec{y}$, which we call the PEE thread, with the density of the threads determined by the boundary PEE structure. In the AdS bulk, the set of all the PEE threads forms a continuous ``network'', which we call the PEE network.
	
In this paper, we show that the density of the PEE threads passing through any bulk point is exactly $1/(4G)$. 
Based on this observation we give a reformulation of the Ryu-Takayanagi (RT) formula for a generic boundary region in general dimensional Poincar\'e AdS space. More explicitly, for any static boundary region $A$, the homologous surface $\Sigma_{A}$ that has the minimal number of intersections with the bulk PEE network is exactly the RT surface of $A$, and the minimal number of intersections reproduces the holographic entanglement entropy. The reconstruction for the area of bulk geometric quantities by counting the number of intersections with the bulk PEE network applies to generic bulk geometric quantities. Interestingly, this reconstruction indicates a pure geometric statement, which is exactly the so-called \emph{Crofton formula} in Poincar\'e AdS.
\end{abstract}

\section{Introduction}
	The AdS/CFT correspondence \cite{Maldacena:1997re,Gubser:1998bc,Witten:1998qj} states that the quantum theory of gravity in asymptotic AdS$_{d+1}$ spacetime is equivalent to a certain CFT$_d$ on the asymptotic boundary. 
	This provides a window into understanding both the classical and quantum aspects of gravitational theories based on the information in the boundary CFT, using the dictionary of the correspondence. 
	Among the studies of the correspondence, holographic entanglement entropy plays a central role, as it relates boundary quantum entanglement structure to the geometry of spacetime. 
	Several important developments have been made along this line \cite{Ryu:2006bv,Hubeny:2007xt,Lewkowycz:2013nqa,Faulkner:2013ana,Faulkner:2013ica,Engelhardt:2014gca,Dong:2013qoa,Dong:2016hjy}.
	These achievements began with the Ryu-Takayanagi (RT) formula \cite{Ryu:2006bv} which links the entanglement entropy of any boundary region and the area of the bulk minimal surface homologous to that boundary region.
	This proposal was later refined into the covariant version \cite{Hubeny:2007xt,Dong:2016hjy} and the version including quantum corrections \cite{Lewkowycz:2013nqa,Faulkner:2013ana,Engelhardt:2014gca,Dong:2013qoa}.
	For more recent developments on the bulk reconstruction inspired by holographic study of quantum entanglement, one consults the following review papers \cite{Bousso:2022ntt,Faulkner:2022mlp}. 
	
	The possibility to reconstruct the bulk geometry from the entanglement structure of the boundary field theory was soon realized after the RT formula was proposed, see \cite{VanRaamsdonk:2009ar,VanRaamsdonk:2010pw} for the earliest discussions. 
	In this paper, we will focus on the explicit reconstruction for the area of bulk geometric quantities in terms of boundary entanglement structure measures. 
	So far, several approaches have been explored for this goal. 
	For example, the reconstruction of certain bulk curves via the \textit{differential entropy} \cite{Balasubramanian:2013lsa,Headrick:2014eia,Czech:2014wka,Czech:2014ppa,Czech:2015qta,Czech:2015kbp} by studying the geodesics tangent to the curve, the reformulation of the RT formula as the maximal flux of the \textit{bit threads} in AdS space out of the region \cite{Freedman:2016zud,Headrick:2020gyq,Headrick:2025awv,Headrick:2022nbe,Harper:2018sdd,Harper:2019lff,Agon:2018lwq,Agon:2021tia,Rolph:2021hgz,Lin:2020yzf,Du:2024xoz}, and the simulation of the AdS space based on the \text{tensor networks} \cite{Swingle:2009bg,evenbly2011tensor,Haegeman:2011uy,Swingle:2012wq,Qi:2013caa,Pastawski:2015qua,Hayden:2016cfa,Bhattacharyya:2016hbx,Bhattacharyya:2017aly} where the RT surface is interpreted as the homologous path in the network with the minimal number of cuts.
	See \cite{Chen:2021lnq} for a detailed review on the above approaches. 
	
	The geometry reconstruction program remains far from complete. 
	The differential entropy scheme works effectively for reconstructing specific curves in AdS$_3$, but becomes intractably complicated in higher dimensions \cite{Balasubramanian:2018uus}. 
	On the other hand, the bit threads scheme works well for static configurations in general dimensions, but it is unclear how to reconstruct geometric quantities beyond the RT surfaces using bit threads. 
	Furthermore, the bit thread configuration depends on the chosen boundary regions, and is highly degenerate even for a given region. 
	Therefore, it is not clear what can be learned from an explicit bit thread configuration beyond holographic entanglement entropy.
	Tensor networks are toy models that can reproduce the key features of AdS/CFT, including the AdS background geometry \cite{Nozaki:2012zj}, the RT formula for holographic entanglement entropy \cite{Swingle:2009bg,Hayden:2016cfa}, and the quantum error correction of holography \cite{Pastawski:2015qua}, among others. 
	However, extending the simulation of AdS/CFT via tensor networks to higher dimensions and time-dependent configurations is subtle and challenging. Additionally, the interpretation of geometric quantities beyond the RT surface in terms of tensor network structure is rarely explored.
	
	Inspired by these approaches, we propose a framework to reconstruct all bulk geometric quantities based on a new measure of entanglement, the partial entanglement entropy (PEE) \cite{Wen:2018whg,Wen:2019iyq,Wen:2020ech,Han:2019scu,Han:2021ycp}.
	See \cite{Kudler-Flam:2019oru,Wen:2018mev,Abt:2018ywl,Rolph:2021nan,Gong:2023vuh,Lin:2022aqf,Lin:2021hqs,Lin:2023jah,Lin:2023hzs,Lin:2023orb} for discussions on the relation between PEE and the above three approaches, and see also \cite{Wen:2021qgx,Camargo:2022mme,Wen:2022jxr,Basu:2022nyl,Basu:2023wmv,Lin:2023ajt} for the reconstruction of the entanglement wedge cross-section in various scenarios based on PEE. 
	In this paper, we focus on the case of a time slice in general dimensional Poincar\'e AdS, where our scheme clearly reconstructs the area of a generic geometric quantity (or any co-dimension two surface) in terms of the boundary PEE structure.

	{
	This paper is based on the geometric scheme in \cite{Lin:2023rbd}, where the geometric counterpart of PEE, namely PEE threads which are geodesics in the time slice, is introduced.
	Bit-thread configurations and entanglement entropies for spherical symmetrical regions are recovered by PEE threads.
	A key observation in \cite{Lin:2023rbd} from AdS$_3$ is that the contribution of a PEE to holographic entanglement entropy is weighted by the number of intersections of its PEE thread with the RT surface.
	In this paper, we provide a proof of this statement in Poincar\'e AdS, and reformulate the area of RT surface, i.e. holographic entanglement entropy, as a double integral.
	And we show that our reformulation is nothing but the \emph{Crofton formula}, and shall be able to extend to general spacetimes.
	}

	In section \ref{sec2}, we will briefly review the PEE in the vacuum state of a holographic CFT, and the setup \cite{Lin:2023rbd} for its geometrization as the PEE threads in the AdS bulk. Also we will go through the calculations \cite{Lin:2023rbd} that count the number of intersections between the RT surfaces of spherical regions and the bulk PEE network, which is very useful for our later discussions. In section \ref{sec3}, we compute the number of intersections between the RT surfaces for generic boundary regions and the bulk PEE network. We show that for any given boundary region, the RT surface is reproduced by identifying the homologous surface that has the minimal number of intersections with the bulk PEE network. Furthermore, this minimal number of intersections is exactly the holographic entanglement entropy given by the RT formula. This gives a reformulation of the RT formula, which is quite similar to the calculation of the entanglement entropy in tensor network models. In section \ref{sec4}, we use the PEE threads to reconstruct a generic bulk co-dimension two surface by counting the number of intersections between the surface and the PEE network. In section \ref{sec5}, we show that our reconstruction scheme is equivalent to a pure mathematical statement, which is precisely the well-known \textit{Crofton formula}. In the last section, we give a summary on our results and give some possible interesting future directions.

	\section{The partial entanglement network } \label{sec2}
	In this section, {we first introduce the basic aspects of PEE. Then we review the main results of \cite{Lin:2023rbd}, which geometrizes the two-point PEE structure into bulk geodesics (namely the PEE threads), and propose to weight the PEE threads differently when conputing holographic entanglement entroy. These results set the stage for our next section, where a reformulation of the RT formula in Poincar\'e AdS will be provided.}

	\subsection{Partial entanglement entropy}
	
	The PEE $\mathcal{I}(A,B)$ is a special measure of two-body correlation between two non-overlapping regions $A$ and $B$ \cite{Wen:2018whg,Wen:2019iyq,Han:2019scu,Han:2021ycp}.
	Besides all the physical properties that are satisfied by mutual information $I(A,B)$, PEE possesses an exclusive property of additivity \cite{Wen:2019iyq,Chen:2014}. More explicitly, assuming that $A$, $B$ and $C$ are three non-overlapping regions, the physical requirements for the PEE are classified in the following
	\begin{enumerate}
		\item
		\textbf{Additivity:} $\mathcal{I}(A,B\cup C)=\mathcal{I}(A,B)+\mathcal{I}(A,C)$;
		
		\item
		\textbf{Permutation symmetry:} $\mathcal{I}(A,B)=\mathcal{I}(B,A)$;
		
		\item
		\textbf{Normalization:} $\mathcal{I}(A,B)|_{B\to \bar{A}}=S_{A}$;
		\item
		\textbf{Positivity:} $\mathcal{I}(A,B)>0$;
		\item
		\textbf{Upper bounded:} $\mathcal{I}(A,B)\leq \text{min}\{S_{A},S_{B}\}$;
		\item
		$\mathcal{I}(A,B)$ should be \textbf{invariant under local unitary transformations} inside $A$ or $B$;
		\item
		\textbf{Symmetry:} for any symmetry transformation $\mathcal T$ under which $\mathcal T A = A'$ and $\mathcal T B = B'$, we have $\mathcal{I}(A,B) = \mathcal{I}(A',B')$.
	\end{enumerate}
	As was shown in \cite{Wen:2019iyq,Casini:2008wt}, the above requirements have a unique solution for states with Poincar\'e symmetry. Furthermore, for the vacuum state of a CFT on a plane, the formula of the solution is totally determined by the above requirements.
	
	According to the properties of additivity and permutation symmetry, the PEE structure is fully described by two-point PEEs $\mathcal I(\vec x,\vec y)$, and $\mathcal I(A,B)$ can be written as a double integral over $A$ and $B$
	\begin{align}
		\mathcal{I}(A,B)=\int_A\D\sigma_{\vec x}\int_{B}\D\sigma_{\vec y}~ \mathcal I(\vec x,\vec y)\,.
	\end{align}
	where $\sigma_{\vec x,\vec y}$ are the infinitesimal area element at $\vec x$ and $\vec y$, and the two-point PEE $\mathcal I(\vec x,\vec y)$ in vacuum CFT$_d$ is given by
	\begin{equation}\label{eq:2p_PEE}
		\mathcal{I}(\vec x,\vec y)=\frac{c}{6}\frac{2^{d-1}(d-1)}{\Omega_{d-2}|\vec x-\vec y|^{2(d-1)}},
	\end{equation}
	where $\Omega_{d-2}={2 \pi^{\frac{d-1}{2}}}/{\Gamma\left(\frac{d-1}{2}\right)}$ is the area of $(d-2)$-dimensional unit sphere. One can either derive the above formula for two-point PEE via the solution to all the physical requirements \cite{Casini:2008wt,Wen:2019iyq}, or use the so-called additive-linear-combination (ALC) proposal in quasi-one-dimensional system to construct PEE \cite{Han:2019scu}. See \cite{Lin:2023rbd} for the details about the derivation of \eqref{eq:2p_PEE}. Although the PEE structure \eqref{eq:2p_PEE} may not capture all the information of the entanglement structure in a CFT, we will see that it is enough to reconstruct the geometric quantities at order $\mathcal{O}(c)$ in the gravity side of AdS/CFT.
	
	The normalization property of the PEE $\mathcal{I}(A,B)|_{B\to \bar{A}}=S_{A}$ tells us how to approach the entanglement entropy from PEE,
	\begin{equation}\label{eq:norm}
		S_{A}=\int_A\D\sigma_{\vec x}\int_{\bar A-\epsilon}\D\sigma_{\vec y}~ \mathcal I(\vec x,\vec y),
	\end{equation}
	where $A\cup \bar{A}$ makes up a pure state and $\epsilon$ represents a regularization cutoff. Note that, the requirement of normalization is quite subtle as it is an equality between two divergent quantities which are normalized in different schemes, see \cite{Han:2019scu} for more discussions on this requirement. {We should keep in mind that, we only impose the normalization requirement to spherical regions for the following reasons. Firstly, imposing this reqruirement {to spherical regions} is enough to determine the two-point PEE structure \eqref{eq:2p_PEE}. Secondly, if we futher impose this requirement to more generic regions, the solution does not exist. In other words, the entanglement entropy for non-spherical regions can not be reproduced by \eqref{eq:2p_PEE} and \eqref{eq:norm} \footnote{In \cite{Bueno:2015rda,Bueno:2015qya,Bueno:2019mex,Bueno:2021fxb}, the entanglement entropies for various shapes of connected regions for the vacuum state of several CFTs have been carried out based on \eqref{eq:2p_PEE} and \eqref{eq:norm}. {Compared with the results derived from the RT formula, these results have a similar formula, but the coefficents at each order are different.} In these papers, the authors studied the mutual information that satisfies additivity (EMI), which coincides with the PEE \cite{Wen:2019iyq} in these scenarios. See \cite{Wen:2019iyq,Han:2019scu} for discussion on the relationship between the PEE and the EMI.}. One can understand this more clearly in the case of one dimensional lattice circle, where the spherical regions are single intervals. It was pointed out in \cite{Wen:2019iyq} that, the number of single intervals equals to the number of two-point PEEs, hence the number of equations \eqref{eq:norm} equals the number of two-points PEEs. Thirdly, as implied in \cite{Lin:2023rbd}, if we want to reproduce the RT formula for multi-intervals, we need to give up this requirement. Later we will see that, the holographic entanglement $S_{A}$ for non-spherical regions can be reproduced from the two-point PEE structure by including contributions from other two-point PEEs, except those represented by $\mathcal{I}(A,\bar{A})$. }

	\subsection{PEE threads and the partial entanglement network}

	In AdS/CFT, we introduced a scheme to geometrize the PEEs in \cite{Lin:2023rbd}, where the boundary two-point PEEs $\mathcal{I}(\vec x,\vec y)$ are represented by the bulk geodesics connecting  two boundary points $\{ \vec x, \vec y\}$, which we call the PEE threads \cite{Lin:2023rbd}. This geometrization looks similar to the bit thread configurations. Nevertheless they are quite different objects, for example the PEE thread configuration is totally determined by the boundary state and they intersect with each other, while the bit thread configuration are highly degenerate and bit threads do not intersect with each other. 
	We only consider the vacuum state of the CFT$_{d}$ and a static time slice in AdS$_{d+1}$. 
	The PEE threads emanating from any point $\vec x$ can be represented by a divergenceless vector field $V_{\vec x}^{\mu}=|V_{\vec x}|\tau^{\mu}$, where $\tau^{\mu}$ is the unit vector tangent to the geodesics emanating from $\vec x$. 
	The norm $|V_{\vec x}|$ characterizes the density of the threads, which is determined by the requirement that,
	\begin{itemize}
		\item \textit{setup: the number of PEE threads connecting any two boundary non-overlapping regions (or boundary points) $A$ and $B$ should match the PEE $\mathcal{I}(A,B)$}.
	\end{itemize}  
	Note that, using vector fields to describe PEE threads endows orientation to the threads, which is indeed redundant information. We stress that the PEE threads are un-oriented lines, and instead of computing the “flux” of the vector field passing through any bulk surface, we will calculate the number of intersections between the surface and the PEE threads. 
	Unlike \cite{Lin:2023rbd}, in this paper we avoid using the words “PEE flux” to prevent unnecessary confusion.
	
\begin{figure}
		\centering
		\begin{tikzpicture}[scale=0.6,
 y={(0:10mm)},x={(-120:5mm)},z={(90:10mm)}]
 % isometric view
 ]
\begin{scope}[canvas is xy plane at z=0]
\def\numsite{4}
\draw[step=2,lightgray,very thin] (-\numsite,-\numsite) grid (\numsite,\numsite);
\draw[black,thick] (-\numsite,-\numsite) rectangle (\numsite,\numsite);
\end{scope}
\foreach \x in {-3,-1,...,3}
% \foreach \y in {-3,-1,...,3}
% \foreach \z in {-3,-1,...,3}
% \foreach \w in {-3,-1,...,3}
{\filldraw[LouisOrange,canvas is xy plane at z=0] (\x,\x) circle (2pt);
\filldraw[LouisOrange,canvas is xy plane at z=0] (\x,-\x) circle (2pt);
}

% \foreach \x in {-3,-1,1}
% \foreach \w in {-3,,...,3}
% \def\circlewidth{0.4*(4/sqrt((\w-\x)^2+(\w-\x)^2))+0.2pt}
% \draw[gray,line width=\circlewidth]{[dashed,rotate around z=180+atan((\x-\w)/(\x-\w)),canvas is xz plane at y=-(\x-\x*(\x-\w)/(\x-\w))*cos(atan((\x-\w)/(\x-\w)))]({(\y-\x*(\y-\w)/(\x-\w))*sin(180+atan((\x-\w)/(\x-\w)))-(\x)/cos(atan((\x-\w)/(\x-\w)))},0) arc(0:180:{sqrt((\w-\x)^2+(\w-\x)^2)/2})};

\foreach \x in {-3,-1,...,3}
% \foreach \y in {-3,-1,...,3}
\foreach \z in {-3,-1,...,3}
% \foreach \w in {-3,-1,...,3}
{\def\y{\x}
\def\w{\z}
\def\circlewidth{0.4*(4/sqrt((\z-\x)^2+(\w-\y)^2))+0.2pt}
\pgfmathparse{\y>\w &&\x==\z?int(1):int(0)}
\let\r\pgfmathresult
\pgfmathparse{\y==\w &&\x>\z?int(1):int(0)}
\let\s\pgfmathresult
\pgfmathparse{\y<\w &&\x<\z?int(1):int(0)}
\let\t\pgfmathresult
\pgfmathparse{\y<\w &&\x>\z?int(1):int(0)}
\let\o\pgfmathresult
\ifnum\r=1
% \draw[gray,line width=\circlewidth]{[canvas is yz plane at x=\x]
% (\y,0)arc (0:180:{sqrt((\y-\w)^2)/2})};
\else \ifnum\s=1
% \draw[gray,line width=\circlewidth]{[canvas is xz plane at y=\y]
% (\x,0)arc (0:180:{sqrt((\x-\z)^2)/2})};
\else \ifnum\t=1
\def\y{-\x}
\def\w{-\z}
\draw[name path=threads1,gray,line width=\circlewidth]{[dashed,rotate around z=180+atan((\y-\w)/(\x-\z)),canvas is xz plane at y=-(\y-\x*(\y-\w)/(\x-\z))*cos(atan((\y-\w)/(\x-\z)))]({(\y-\x*(\y-\w)/(\x-\z))*sin(180+atan((\y-\w)/(\x-\z)))-(\x)/cos(atan((\y-\w)/(\x-\z)))},0) arc(0:180:{sqrt((\z-\x)^2+(\w-\y)^2)/2})};
\def\y{\x}
\def\w{\z}
\draw[name path=threads2,gray,line width=\circlewidth]{[dashed,rotate around z=180+atan((\y-\w)/(\x-\z)),canvas is xz plane at y=-(\y-\x*(\y-\w)/(\x-\z))*cos(atan((\y-\w)/(\x-\z)))]({(\y-\x*(\y-\w)/(\x-\z))*sin(180+atan((\y-\w)/(\x-\z)))-(\x)/cos(atan((\y-\w)/(\x-\z)))},0) arc(0:180:{sqrt((\z-\x)^2+(\w-\y)^2)/2})};
% \draw[name intersections={of=threads1 and threads2, name=i,total=\t}][fill=red]{\foreach \s in {1,...,\t} (i-\s)circle(2pt)};
\else \ifnum\o=1
% \def\y{-\x}
% \def\w{-\z}
% \draw[gray,line width=\circlewidth]{[dashed,rotate around z=180+atan((\y-\w)/(\x-\z)),canvas is xz plane at y=-(\y-\x*(\y-\w)/(\x-\z))*cos(atan((\y-\w)/(\x-\z)))]({(\y-\x*(\y-\w)/(\x-\z))*sin(180+atan((\y-\w)/(\x-\z)))-(\x)/cos(atan((\y-\w)/(\x-\z)))},0) arc(180:0:{sqrt((\z-\x)^2+(\w-\y)^2)/2})};
\fi
\fi
\fi
\fi
}
\draw[fill=red](0,0,{sqrt(2)})circle(2pt);
\draw[fill=red](0,0,{sqrt(6)})circle(2pt);
\draw[fill=red](0,0,{sqrt(18)})circle(2pt);
\end{tikzpicture}
		\includegraphics[scale=0.45]{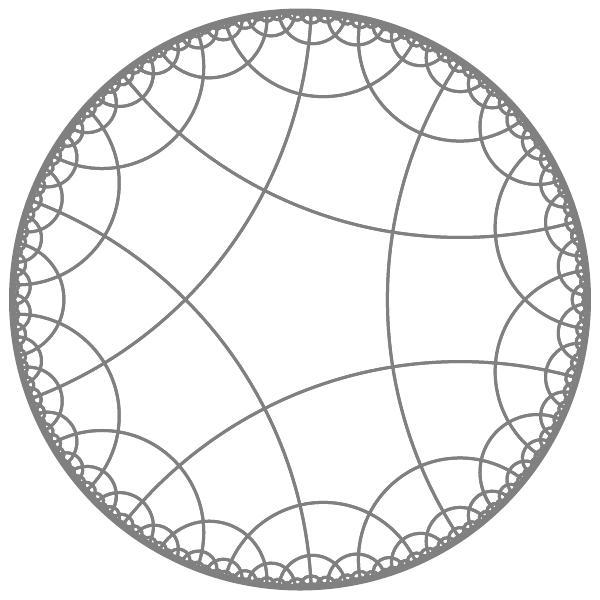}
		\caption{Visualizations of PEE threads (gray) on a time slice of Poincar\'e AdS$_4$ and global AdS$_3$, where the boundary sites (on the plane) are discretized.
			Upper: the PEE threads with larger radii are denoted with thinner curves, whose thickness reflects the density of the PEE threads.
			Bottom: with appropriate discretizations, they form a hyperbolic tiling.
		}
		\label{fig:PT_slice}
	\end{figure}
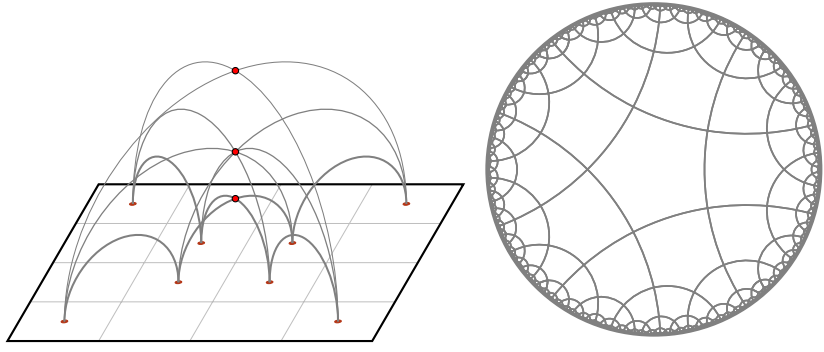

	In summary, given the PEE structure $\mathcal{I}(\vec x, \vec y)$ of the boundary CFT and the metric of the dual spacetime, we get a network of the PEE threads in the AdS bulk consisting of all the bulk geodesics on a time slice  (see Fig.\ \ref{fig:PT_slice} for examples) and the density of the geodesics is totally determined by the boundary PEE structure. We call it the \textit{partial entanglement (entropy) network}, or the \textit{PEE network} for short.
	
	We briefly review the derivation of the PEE vector flow $V_{\vec x}^\mu$ and its connection with the configuration of bit threads \cite{Lin:2023rbd} when $A$ is a spherical region.
	Due to the translation symmetry, it is sufficient to derive the PEE threads emanating from the origin $O$, $V_O^\mu\equiv V_{\vec x=0}^\mu$.
	We work in poincar\'e AdS$_{d+1}$ with unit AdS radius,
	\begin{align}
		\D s^{2}=\frac{-\D t^2+\D z^2+\D r^2+r^2\D\Omega_{d-2}^{2}}{z^2}\,.
	\end{align}
In higher dimensions, due to the rotational symmetry, we restrict to a 2-dimensional slice with $\phi_i=0$. 
	Since the PEE threads are just the bulk geodesics anchored on $O$, the vector field $V_O^{\mu}(Q)$ is tangent to these geodesics, such that 
	\begin{equation}
		V_{O}^\mu(Q)=
		|V_{O}(Q)|\tau_{O}^\mu(Q),
	\end{equation}
	where
	\begin{equation}
		\tau_{O}^\mu(Q)=\frac{2 z r}{ r^2+ z^2}\left(
		z,\frac{ z^2- r^2}{2 r}
		\right),
	\end{equation}
	is the unit vector tangent to the geodesics emanating from $O$, and $z,~r$ are the coordinates of the bulk point $Q$ on the $\phi_i=0$ slice.
	The norm $|V_{O}(Q)|$ is then settled down by the requirement that, the number of PEE threads from $O$ to any boundary region $\D y$ should match the PEE $\mathcal{I}(0, \D  y)$, where $y$ is the radial coordinate on the boundary.
	
	\begin{figure}[h]
		\centering
		\begin{tikzpicture}[scale=1]
			\coordinate [label=below:$O$] (A) at (0,-0.1);
			\draw[very thick] (-3,0) -- (5,0);
			\draw[ultra thick,LouisBlue] (-2.5,0) arc (180:0:2.5) ;
			\draw[LouisOrange,middlearrow={0.66},dashed,thick](0,0) arc(180:0:2.1);
			\filldraw [LouisOrange] (53.4704:2.5) circle (1.5pt);
			\draw[black,thick](0,0)--(53.4704:2.5);
			\draw[LouisOrange,middlearrow={0.77},dashed,thick](0,0) arc(180:0:1.5);
			\filldraw [LouisOrange] (33.5573:2.5) circle (1.5pt);
			\draw [] (3.6,0) node[label={[scale=1]$\mathrm d y$}]{};
			\draw [] (33:2.6) node[label={[scale=1]$\mathrm d \Sigma$}]{};
			\filldraw[black] (0,0) circle (1.5pt);
			\draw[] (-2.5,-0.3) node {$-\ell$};
			\draw[] (2.4,-0.3) node {$\ell$};
			\draw[] (4.2,-0.3) node {$y$};
			\draw[] (60:2.7) node {$Q$};
			\draw[] (0.5,0.3) node {$\theta$};
			\filldraw[black] (3,0) circle (1.5pt);
			\filldraw[black] (4.2,0) circle (1.5pt);
		\end{tikzpicture}
		\caption{This figure is extracted from \cite{Lin:2023rbd}. Here $\D\Sigma$ is an infinitesimal area element at $Q$, and the blue circle is the reference RT surface $r^2+z^2=\ell^2$. The PEE threads emanating from $O$ and passing through $\D\Sigma$ will anchor on a boundary region $\D y$. There exists an one-to-one mapping between any point $(\ell\cos\theta,\ell\cos\theta)$ on $\D\Sigma$ and the point $y=\ell/\cos\theta$ in the $\D y$ region. This pair of points are connected by a PEE thread emanated from $O$.}
		\label{fig:pt-v}
	\end{figure}
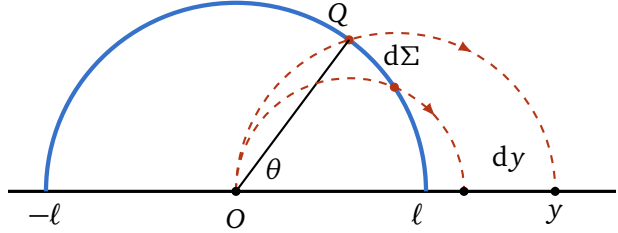
	
	More explicitly, let us consider the coordinate $Q=(\bar{r},\bar{z})=(\ell \cos\theta,\ell \sin\theta)$ and a reference RT surface $\Sigma$ passing through $Q$ with the radius $\ell=\sqrt{ r^2+ z^2}$ (see Fig.\ \ref{fig:pt-v}). The number of the PEE threads $V_{O}^{\mu}$ passing through an area element $\D\Sigma$ at $Q$ on $\Sigma$ is given by
	\begin{equation}
		\begin{aligned}
			\mathcal{F}(V_{O}^{\mu},\D\Sigma)
			=\D\theta\D\Omega_{d-2}\sqrt{h_{\theta\theta}}|V_O(\theta)|\cot^{d-2}\theta\sin\theta,
		\end{aligned}
	\end{equation}
	where $\theta=\arctan\frac{ z}{ r}$,  $h_{\theta\theta}=1/\sin^2\theta$ is the $\theta\theta$-component of the induced metric on $\Sigma$ and $ n_{\Sigma}^\mu(Q)=\ell\sin\theta(\cos\theta,\sin\theta)$ is the unit normal vector on $\Sigma$. 
	We have used $\sqrt{h}=\frac{\bar r^{d-2}}{\bar z^{d-2}}\sqrt{h_{\theta\theta}}$.
	On the other hand, the PEE $\mathcal{I}(O, \D y)$ is given by $ y^{d-2}\mathcal{I}(0,y)\Omega_{d-2}\D y$. Then we solve the requirement 
	\begin{equation}
		\mathcal{F}(V_{O}^{\mu},\D\Sigma)=\D y~ y^{d-2}\mathcal{I}(0,y)\Omega_{d-2},\quad \ y=\ell/\cos\theta
	\end{equation}
	to get the norm of $V_{O}(Q)$,
	\begin{equation}
		|V_{O}(Q)|=\frac{1}{4G}\frac{2^{d-1}(d-1)}{\Omega_{d-2}}\frac{ z^{d-1}}{( r^2+ z^2)^{d-1}},
	\end{equation}
	where we have replaced $(\bar r, \bar z)$ with $(r, z)$. In the above derivation we used the mapping $y=\ell/\cos\theta$ since $\D \Sigma$ determines the $\D y$ region on the boundary following the PEE threads. Finally, we obtain the PEE threads flow on the $\phi_i=0$ slice
	\begin{equation}\label{pv}
		V_{O}(Q)=\frac{2^d  z^d}{4G}\frac{(d-1)}{\Omega_{d-2}}\frac{ r}{( r^2+ z^2)^d}
		\left(
		z,\frac{ z^2- r^2}{2 r},0,\cdots
		\right)\,.
	\end{equation}
	Due to the translation symmetry, the vector flow from any boundary point $\vec x$ is identical to \eqref{pv} up to a translation.
	For example, in AdS$_3$, the PEE thread flow emanating from $r=r_0$ can be obtained by replacing $r$ with $ r-r_0$ in \eqref{pv} and
	\begin{equation}
		V_{r_0}^\mu=\frac{1}{4G}\frac{2z^2(r-r_0)}{(( r-r_0)^2+ z^2)^2}\left(z,\frac{ z^2-( r-r_0)^2}{2( r-r_0)}\right).
	\end{equation}
	
	Now we sum up the PEE threads emanating from all the points inside a spherical region $A=\{\textbf x||\textbf x|<R\}$, to get a new vector flow\footnote{In \cite{Lin:2023rbd} it was pointed out that this vector field is exactly the bit thread \cite{Freedman:2016zud} configuration for $A$ constructed in \cite{Agon:2018lwq}. Hence, this bit thread flow is just a superposition of the PEE flow emanating from all the points in the region under consideration.}
	\begin{align}\label{VA}
		V_A^\mu&
		=\int_A\D^{d-1}\vec x V_{\vec x}^\mu =\frac{1}{4G}\left(\frac{2 R z}{\sqrt{\left(R^2+r^2+z^2\right)^2-4 R^2 r^2}}\right)^{d}\left(\frac{r z}{R}, \frac{R^2-r^2+z^2}{2 R}\right).
	\end{align}
	When performing the above integration, we have defined the orientation of the threads to point from $A$ to $\bar{A}$. One can check that the vector field $V_A^\mu$ at any point $Q$ on the RT surface $\mathcal{E}_{A}$ of $A$ is just \cite{Lin:2023rbd}
	\begin{align}\label{fv}
		\int_A V_{\vec x}^{\mu}(Q)\D^{d-1}\vec x=\frac{1}{4G}n^{\mu}(Q), \qquad Q\in \mathcal{E}_{A}\,,
	\end{align}
	where $n^\mu(Q)$ is the outward normal unit vector on the RT surface $\mathcal{E}_{A}$. This result is independent of the radius and position of $A$. 
	Note that, for spherical regions $A$, all the PEE threads connecting $A$ and $\bar{A}$ intersect with $\mathcal{E}_{A}$ once, hence the ``flux'' of PEE threads \eqref{fv} passing through $\mathcal{E}_{A}$ equals to the number of intersections between $\mathcal{E}_{A}$ and the PEE network. In summary, we arrive at the following important statement:
	\begin{itemize}
		\item \textit{Statement 1: the density of intersections between the bulk PEE network and the RT surface $\mathcal{E}_{A}$ for any spherical region (or single interval) $A$ is given by the constant $\frac{1}{4G}$ everywhere on $\mathcal{E}_{A}$.}
	\end{itemize}
	Later, we denote the number of intersections between any bulk co-dimension-two surface $\Sigma$ and the PEE network as $\mathcal{N}(\Sigma)$. So we have that, for spherical regions $A$
	\begin{align}
		\mathcal{N}(\mathcal{E}_{A})=\frac{\text{Area}[\mathcal{E}_{A}]}{4G}\,,
	\end{align}
	which coincides with the holographic entanglement entropy given by the RT formula.
	
	On the other hand, one should note that \eqref{fv} does not hold for non-spherical regions, see \cite{Lin:2023rbd} for a calculation of \eqref{fv} when $A$ is a strip. 
	
	\subsection{The weight of the PEE threads}
	Also in \cite{Lin:2023rbd} the authors explored cases where $A$ goes beyond spherical regions. Let us denote $\Sigma_A$ as an arbitrary surface homologous to $A$, and $\mathcal E_A$ as the RT surface of $A$, hence
	\begin{align}
		\mathrm{Area}[\mathcal E_A]=\min_{\Sigma_{A}}{\mathrm{Area}[\Sigma_A]}.
	\end{align}
	If we naively apply the normalization property \eqref{eq:norm} to any region $A$, then we can calculate the entanglement entropy $S_A$ based on the bulk PEE network in the following steps: 1) we define the orientation of the threads connecting $A$ and $\bar A$ to point from $A$ to $\bar A$; 2) we integrate the vector fields as in \eqref{VA} to get the vector field $V_A^{\mu}$; 3) we counts the number of the PEE threads connecting $A$ and $\bar A$ by calculating the following integration, 
	\begin{equation}\label{eq:pee_prop}
		{\mathcal{I}(A,\bar{A})}=\int_{\Sigma_A}\D\Sigma_A\sqrt{h}V_{A}^\mu n_\mu.
	\end{equation}
	Naively, we expect that \eqref{eq:pee_prop} should reproduce the same entanglement entropy as the RT formula for an arbitrary region $A$. Nevertheless, this is not true for non-spherical regions, see \cite{Lin:2023rbd} for explicit analysis when $A$ is a strip region. {This implies that the normalization property only applies to spherical regions in Poincar\'e AdS, and \eqref{eq:norm} should be modified to reproduce the holographic entanglement entropy for more generic boundary regions.} 
	
	\begin{figure}
		\centering
		\begin{tikzpicture}[scale=0.7]
			\clip (-6,-1) rectangle (6,4);
			% \filldraw[LouisColor2](-7,0) rectangle (7,6);
			\filldraw[LouisBlue!10](-4,0) arc (180:0:1.5);
			\filldraw[LouisBlue!10](1,0) arc (180:0:1.5);
			\draw[ultra thick,LouisBlue] (-4,0) arc (180:0:1.5) ;
			\draw[ultra thick,LouisBlue] (1,0) arc (180:0:1.5) ;
			\draw[very thick] (-5.5,0) -- (5.5,0);
			\draw[] (0,-0.4) node {$\bar{A}_1$};
			\draw[] (2.5,-0.4) node {$A_2$};
			\draw[] (-2.5,-0.4) node {$A_1$};
			\draw[] (-5,-0.4) node {$\bar{A}_2$};
			\draw[] (5,-0.4) node {$\bar{A}_2$};
			\draw[ thick,dashed,LouisColor1] (-1.5,0) arc (180:0:1.5) ;
			\draw[ thick,dashed,gray] (0,0) arc (180:0:2.5) ;
			\draw[ thick,dashed,LouisOrange] (-5,0) arc (180:0:1) ;
			\draw[ thick,dashed,LouisColor1] (-5.7,3.2) -- (-5,3.2) ;
			\draw[] (-3.4,3.25) node {$\omega(A_1,A_2)=2$};
			\draw[ thick,dashed,gray] (-5.7,2.) -- (-5,2.) ;
			\draw[] (-3.4,2.05) node {$\omega(\bar{A}_1,\bar{A}_2)=0$};
			% \draw[ thick,dashed,LouisColor2] (-5,0) arc (180:0:1) ;
			\draw[ thick,dashed,LouisOrange] (-5.7,2.6) -- (-5,2.6) ;
			\draw[] (-3.6,2.65) node {$\omega(A,\bar{A})=1$};
			% \draw[] (0,3.5) node {$\alpha<1/2$};
		\end{tikzpicture}
		\caption{This figure is extracted from \cite{Lin:2023rbd}. Here $A=A_1\cup A_2$, $\bar{A}=\bar{A}_1\cup \bar{A}_2$ and $A\bar{A}$ is in the vacuum state of the holographic CFT$_2$. Representative PEE threads (dashed curves) in Poincar\'e AdS$_3$.
			Each PEE thread crosses the disconnected RT surface for different number of times, which represents different weight.
		}
		\label{fig:exp}
	\end{figure}
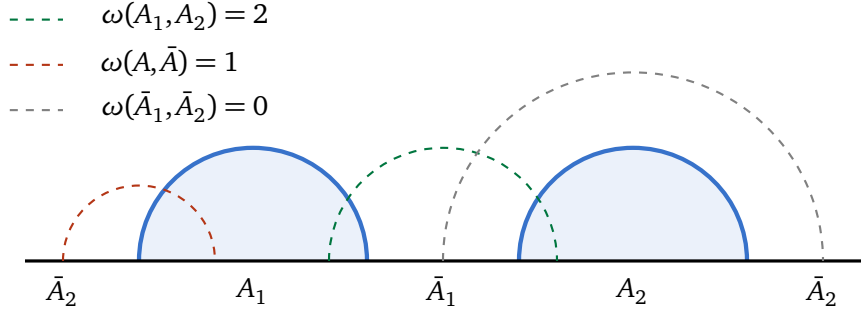
	
	The discussion for the two-interval case in \cite{Lin:2023rbd} in CFT$_2$ provides a clear clue for the modification. Let us consider a two interval region $A=A_1\cup A_2$ whose RT surface is given by the blue curves, see Fig.\ \ref{fig:exp}. 
	If we naively apply \eqref{eq:norm}, we should only count the threads connecting $A_i$ and $\bar{A}_j$, which gives $S_{A}=\mathcal{I}(A_1,\bar{A})+\mathcal{I}(A_2,\bar{A})$. 
	However, the RT formula implies
	\begin{align}\label{RTtwoint}
		S_{A}=&S_{A_1}+S_{A_2}=\mathcal{I}(A_1,A_2\bar{A})+\mathcal{I}(A_2,A_1\bar{A})
		\cr
		=&\mathcal{I}(A_1,\bar{A})+\mathcal{I}(A_2,\bar{A})+2\mathcal{I}(A_1,A_2)\,,
	\end{align}
	where we used \eqref{eq:norm} for $A_1$ and $A_2$ as it applies to single intervals. 
	So, \eqref{RTtwoint} indicates that the RT formula not only counts the threads connecting $A$ and $\bar{A}$, but also doubly counts the threads connecting $A_1$ and $A_2$.
	This absolutely goes beyond \eqref{eq:norm}, but looks reasonable as the threads connecting $A_1$ and $A_2$ intersect with the RT surface twice. 
	
Then one may weight the threads with the number of times they intersect the RT surface. 
For any disjoint multi-interval $A=\cup A_i$ and its complement $\bar{A}=\cup \bar{A}_j$, given the RT surface we can read out the weight for any PEE threads. 
Let us denote the weight for the threads connecting any two sub-intervals $\alpha$ and $\beta$ as $\omega(\alpha,\beta)$.
It has further been checked in \cite{Lin:2023rbd} that the holographic entanglement entropy is given by
	\begin{align}\label{SAmultint}
		S_{A}=\sum_{i,j}\big(\omega(A_i, A_j) \mathcal{I}(A_i&,A_j)+\omega(\bar{A}_i, \bar{A}_j) \mathcal{I}(\bar{A}_i,\bar{A}_j)
		+\omega(A_i, \bar{A}_j) \mathcal{I}(A_i,\bar{A}_j)\big)\,.
	\end{align}
 From bulk PEE network perspective, it is important to note that, \eqref{SAmultint} indeed counts $\mathcal{N}(\mathcal{E}_{A})$, the number of intersections between the RT surface $\mathcal{E}_{A}$ and the bulk PEE network.
	
	\section{Reformulation of the RT formula}\label{sec3}
	One of the main task of this paper is to give a reformulation for the RT formula based on the PEE threads setup. {As we have just reviewed \cite{Lin:2023rbd}, the area of the RT surfaces can be reproduced by counting the number of intersections $\mathcal{N}(\mathcal{E}_{A})$ between the RT surface $\mathcal{E}_{A}$ and the PEE network. Nevertheless, this observation is still far from a reformulation of the RT formula for the following reasons.}
	\begin{itemize}
		\item { Firstly, the discussions in \cite{Lin:2023rbd} were based on the assumption that we already know the RT surface, and no method was provided to identify the RT surface. 
		}
		
		\item {Secondly, the computation for $\mathcal{N}(\mathcal{E}_{A})$ in \cite{Lin:2023rbd} was performed only for the RT surfaces of spherical regions in general dimensions, and multi-intervals in 2-dimensions. In order to give a reformulation for the RT formula, we need to compute the $\mathcal{N}(\mathcal{E}_{A})$ for generic RT surfaces.	}	
	\end{itemize}

	To fulfill our task, our strategy is to prove the following statement, which is the key to give our reformulation of the RT formula.
	\begin{itemize}
		\item \textit{Statement 2: {in Poincar\'e AdS space,} given a generic boundary region $A$ and an arbitrary surface $\Sigma_{A}$ homologous to $A$, the density of intersections between $\Sigma_{A}$ and the bulk PEE network everywhere on $\Sigma_{A}$ is also given by $\frac{1}{4G}$}.
	\end{itemize}
	Provided that the above statement is true, then the total number of intersections is just given by
	\begin{align}
		\mathcal{N}(\Sigma_{A})=\frac{\mathrm{Area}[\Sigma_A]}{4G}\,.
	\end{align}
	And the RT surface $\mathcal{E}_{A}$ can be identified by finding the homologous surface that minimizes the number of intersections $\mathcal{N}(\Sigma_A)$ among all possible $\Sigma_{A}$. Finally, provided the \textit{statement 2} the RT formula is reformulated by
	\begin{align}\label{reformulationRT}
		S_{A}=\min_{\Sigma_{A}}\mathcal{N}(\Sigma_{A})=\frac{\mathrm{Area}[\mathcal{E}_A]}{4G}\,.
	\end{align}
	In other words, given an arbitrary boundary region $A$, the RT surface is the $\Sigma_A$ that has the minimal number of intersections with the PEE network.
	This number is exactly the holographic entanglement entropy $S_{A}$ calculated by the RT formula. The remaining task in this section is to prove the \textit{statement 2}.

	Before we prove the \textit{statement 2}, we give two explicit formulas to calculate the number of intersections $\mathcal{N}(\Sigma_{A})$ for a generic homologous surface. In a time slice of Poincar\'e AdS$_{d+1}$ in general dimensions, let us consider an arbitrary boundary region $A$ (connected or disconnected) and an arbitrary co-dimension two surface $\Sigma_{A}$ homologous to $A$. 
	Similarly we define the weight of a PEE thread as the number of times it intersects with the surface $\Sigma_{A}$, and write it as $\omega_{\Sigma_{A}}(\vec x, \vec y)$, where  $\{\vec x, \vec y\}$ are the boundary points where the PEE threads anchor. Then the number of intersections $\mathcal{N}(\Sigma_{A})$ is calculated by
	\begin{align}\label{eq:wpee-proposal0}
		\boxed{
			\mathcal{N}(\Sigma_{A})=\frac12
		\int_{\partial \mathcal{M}} \D^{d-1}\vec x\int_{\partial \mathcal{M}}\D^{d-1}\vec y~\omega_{\Sigma_A}(\vec x,\vec y)\mathcal I(\vec x,\vec y),
		}
	\end{align}
	where the integration domain of $\vec x$ and $\vec y$ is the whole AdS boundary $\partial \mathcal{M}=A\bar{A}$. 
	Here we have considered all the PEE threads and their weights instead of only those connecting $A$ and $\bar{A}$.
	The factor $1/2$ appears as we count both $\mathcal{I}(\vec x, \vec y)$ and $\mathcal{I}(\vec y, \vec x)$ in the integration, hence every PEE thread has been counted twice. Conducting the integration \eqref{eq:wpee-proposal0} is formidable since it is impossible to trace the weight $\omega_{\Sigma_A}(\vec x,\vec y)$ of all the PEE threads. The formula \eqref{eq:wpee-proposal0} will be quite useful when we relate our results to the \textit{Crofton formula}.
	
	On the other hand, we can also count $\mathcal{N}(\Sigma_{A})$ along the homologous surface $\Sigma_{A}$, since locally the density of PEE threads is determined by the vector fields $V_{\vec x}^\mu(Q)$. Consider an area element $\D\Sigma_{A}$ at the point $Q$ and the unit normal vector $n^{\mu}(Q)$ pointing from one side of $\Sigma_{A}$ to the other side, the number of intersections between $\D\Sigma_{A}$ and the PEE threads emanating from $\vec x$ is given by $|V_{\vec x}^\mu(Q) n_\mu(Q)|\D\Sigma_A$\footnote{Here we take the absolute value for $V_{\vec x}^\mu n_\mu$ since locally we are always counting the number of intersections hence any PEE thread passing through $\Sigma_{A}$ should give positive contribution regardless its orientation.}, then the total number of intersections is given by the following integration over $\Sigma_{A}$,
	\begin{equation}\label{eq:wpee-proposal1}
		\boxed{
			\mathcal{N}(\Sigma_{A})=\frac12\int_{\Sigma_A}\D\Sigma_A\sqrt{h}\int_{\partial \mathcal{M}}\D ^{d-1}\vec x|V_{\vec x}^\mu(Q) n_\mu(Q)|\,.
		}
	\end{equation}
	Here $\D\Sigma_A\sqrt{h}$ is the area of an infinitesimal area element on $\Sigma_{A}$, and the coefficient $1/2$ appears because we integrate $\vec x$ over the whole boundary such that the PEE thread connecting any two boundary point is doubly counted.  The equivalence between \eqref{eq:wpee-proposal0} and \eqref{eq:wpee-proposal1} is guaranteed by the divergenceless property of the vector fields $V_{\vec x}^\mu $ \footnote{{Usually it will be formidable to compute \eqref{eq:wpee-proposal0} for a generic $\Sigma_{A}$, as it will be hard to trace the weight for all of the threads. Here we give a simple example where both of \eqref{eq:wpee-proposal1} and \eqref{eq:wpee-proposal0} can be computed, and the results conincide. Let us consider the two-interval case with a disconnected entanglement wedge shown in Fig.\ref{fig:exp}. Given the RT surface $\mathcal{E}_{A}=\mathcal{E}_{A_1}\cup \mathcal{E}_{A_2}$, it is easy to count the weight for each thread, and the computation of \eqref{eq:wpee-proposal0} is exactly given by \eqref{RTtwoint}, which gives $\mathcal{N}(\mathcal{E}_{A})=S_{A_1}+S_{A_2}=\frac{\mathrm{Area}[\mathcal{E}_{A_1}]+\mathrm{Area}[\mathcal{E}_{A_2}]}{4G}$, where we have used the normalization property for single intervals. }
			
	{On the other hand, Since both of $\mathcal{E}_{A_1}$ and $\mathcal{E}_{A_2}$ are RT surfaces for single intervals, the statement 1 applies hence the density of intersections between these two surfaces and the PEE network is given by the constant $\frac{1}{4G}$. Then we also get $\mathcal{N}(\mathcal{E}_{A})=\mathcal{N}(\mathcal{E}_{A_1}\cup \mathcal{E}_{A_2})=\frac{\mathrm{Area}[\mathcal{E}_{A_1}]+\mathrm{Area}[\mathcal{E}_{A_2}]}{4G}$.
	}}.

\begin{figure}
		\centering
		\begin{tikzpicture}[scale=0.5]
			\clip (-7,-1) rectangle (7,6.2);
			\draw[ultra thick,LouisBlue] (-4,0) arc (180:0:4) ;
			\draw[very thick] (-9,0) -- (9,0);
			% \draw[] (0,-0.5) node {$R_\Sigma$};
			\draw[] (0,-0.4) node {$\mathcal{A}$};
			% \filldraw[black] (0,0) circle (2pt);
			\filldraw[black] (-4,0) circle (2pt);
			\filldraw[black] (4,0) circle (2pt);
			\draw[ultra thick,LouisBlue,fill=LouisBlue!10] (-4,0)arc(180:0:4)--cycle ;
			\draw[thick, gray,middlearrow={0.3}] ({2-2*sqrt(3)},0)arc(180:0:{2*sqrt(3)});
			\draw[thick, gray,middlearrow={0.07}] (1.4641,0)arc(180:0:11.4641);
			\draw[thick, gray,middlearrow={0.3}] (-3,0)arc(180:0:3.7);
			\draw[thick, gray,middlearrow={0.25}] (0,0)arc(180:0:4);
			\draw[thick, gray,middlearrow={0.25}] (3,0)arc(0:90:6.5);
			\draw[ultra thick, LouisOrange] (2.23677,3.31615) arc (56:64:4);
			\draw[] (2.5,{2*sqrt(3)+0.8}) node[scale=1,fill=white,inner sep=1pt]{\color{LouisOrange}$\mathrm d\Sigma$};
		\end{tikzpicture}
		\caption{Here we consider an arbitrary bulk area element $\D\Sigma$ (the red area element) in  Poincar\'e AdS$_3$, and the associated the RT surface of the unique spherical region $\mathcal{A}$ in which $\D\Sigma$ is embedded. It is obvious that counting the PEE threads emanating from $\mathcal{A}$ (denoted by gray solid curves) gives the number of intersections $\mathcal{N}(\D\Sigma)$. This result generalizes to AdS space in general dimensions.
		}
		\label{fig:PEE_Sigma}
	\end{figure}
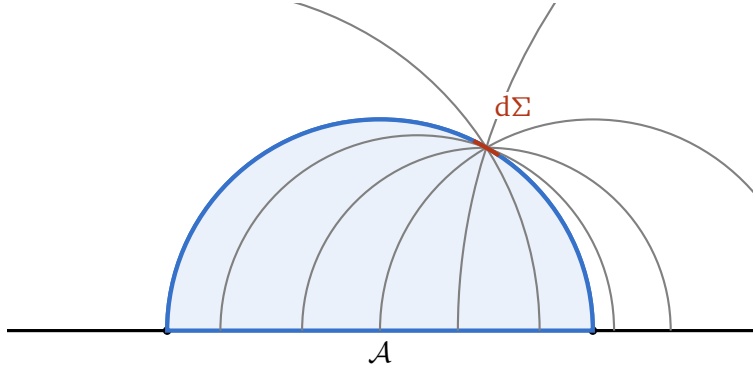

	According to \eqref{eq:wpee-proposal1}, the density of intersections between $\Sigma_{A}$ and the PEE network at any point $Q$ on $\Sigma_{A}$ is just given by
	\begin{align}\label{density}
		\textit{density of intersections}~~=~~\frac12\int_{\partial \mathcal{M}}\D ^{d-1}\vec x|V_{\vec x}^\mu(Q) n_\mu(Q)|\,.
	\end{align}
	Since $\Sigma_{A}$ is an arbitrary homologous surface, {given any bulk point $Q$ and any normal direction $n_\mu(Q)$, if we can calculating the above integration and prove 
	\begin{align}\label{density1}
		\frac12\int_{\partial \mathcal{M}}\D ^{d-1}\vec x|V_{\vec x}^\mu(Q) n_\mu(Q)|=\frac{1}{4G}\,,
	\end{align}
    then the statement 2 follows.}
    {
	For example, in Poincar\'e AdS$_3$ as Fig.\,\ref{fig:PEE_Sigma}, \eqref{density} sums over all the PEE threads that pass through $\Sigma_A$ at $\mathrm{d}\Sigma|_Q$, shown as the gray geodesics in Fig.\,\ref{fig:PEE_Sigma}. The factor $(1/2)$ accounts for the double counting of each geodesic.
	And \eqref{density1} is to say that such a density is a universal constant $1/4G$.
	Detailed explaination and proof of \eqref{density1} go as follows.
	} 
    
    {Let us consider an arbitrary area element $\D\Sigma$ at any point $Q$ in the bulk and the associated normal unit vector $n^{\mu}(Q)$, directly calculating the integration \eqref{density} is complicated, because of the absolute symbol in the integrand. We always start from certain configuration of orientations for all the threads as we use vector fields to describe them in \eqref{density}. Nevertheless, we stress that the PEE threads are un-oriented lines, and the orientation come with the vector field description are redundant information, which are excluded by the absolute value sign. For an arbitrary configuration of the orientations, it is also formidable to compute \eqref{density} due to the absolute value sign. Given an infinitesimal area element in the bulk, the trick of the computation is to find a proper assignment for the orientations of the threads passing through it, such that all the threads point from one same side of the area element to the other side, hence the cancellation due to opposite orientations can be avoided. This allows us to drop the absolute value sign and the number of intersections equals to the flux passing through the area element.}
	
	More explicitly, our strategy is to find the boundary region $\mathcal{B}$ such that, for any $\vec x\in \mathcal{B}$ we have
	\begin{align}\label{propertyB}
		V_{\vec x}^\mu(Q) n_\mu(Q)\geq  0\,,\quad or\quad  V_{\vec x}^\mu(Q) n_\mu(Q)\leq  0\,,\quad \vec x\in \mathcal{B}\,.
	\end{align}
	This means that PEE threads emanating from $\mathcal{B}$ with the initial orientation point from the bulk into the boundary should all pass through $\D\Sigma$ from one side to the other. Then we can count the density of intersections without double counting, i.e.
	\begin{align}
		\textit{density of intersections}~~=~~\int_{\mathcal{B}}\D ^{d-1}\vec x V_{\vec x}^\mu(Q) n_\mu(Q)\,.
	\end{align}
	Note that since every intersection is counted only once, compared with \eqref{density} the $1/2$ factor is gone and the domain of integration becomes $\mathcal{B}$ instead of $\partial \mathcal{M}$. Although the analysis for static spherical regions \cite{Lin:2023rbd} looks quite special, it contains the key ingredient to prove \eqref{density1}.

	In the next we will show that, for any $\D\Sigma$ the corresponding $\mathcal{B}$ is exactly a spherical region. For any $\D\Sigma$, it uniquely determines a boundary spherical region $\mathcal{A}$ whose RT surface $\mathcal{E}_{\mathcal{A}}$ passes through $Q$ and has the same normal vector $n^{\mu}(Q)$ at $Q$ (see Fig. \ref{fig:PEE_Sigma} for an example). 
	In other words $\D\Sigma$ can be embedded into the RT surface of a unique spherical region $\mathcal{A}$. After we determine the spherical region $\mathcal{A}$ for $\D\Sigma$ and its RT surface (a hemisphere) $\mathcal{E}_{\mathcal{A}}$, all the PEE threads can be classified into three classes,
	\begin{enumerate}
		\item $\omega_{\mathcal{E}_{\mathcal{A}}}(\vec x, \vec y)=0$ for $\vec x, \vec y \in \mathcal{A}$;
		\item $\omega_{\mathcal{E}_{\mathcal{A}}}(\vec x, \vec y)=0$ for $\vec x, \vec y \in \bar{\mathcal{A}}$;
		\item $\omega_{\mathcal{E}_{\mathcal{A}}}(\vec x, \vec y)=1$ for $\vec x\in \mathcal{A}, \vec y \in \bar{\mathcal{A}}$ or $\vec x\in \bar{\mathcal{A}}, \vec y \in \mathcal{A}$.
	\end{enumerate}
	We see that only the threads in the third class intersect with $\mathcal{E}_{\mathcal{A}}$ once. 
	In other words, for a specific area element $\D\Sigma$ embedded in $\mathcal{E}_{\mathcal{A}}$, all of the PEE threads passing through $\D\Sigma$ from the inside to the outside are given by the PEE threads emanating from $\mathcal{A}$ with the initial orientation pointing from the boundary to the bulk. This means the spherical region $\mathcal{A}$ determined by $\D\Sigma$ is exactly the region $\mathcal{B}$ which satisfies \eqref{propertyB}. See Fig. \ref{fig:PEE_Sigma} for an explicit illustration. Then using the \textit{statement 1} (or \eqref{fv}) for spherical regions, we arrive at
	\begin{align}
		\frac12\int_{\partial \mathcal{M}}\D ^{d-1}\vec x|V_{\vec x}^\mu(Q) n_\mu(Q)|=\int_{\mathcal{B}}\D ^{d-1}\vec x V_{\vec x}^\mu(Q) n_\mu(Q)=\frac{1}{4G}n^\mu(Q)n_\mu(Q)=\frac{1}{4G}\,,
	\end{align}
	which eventually leads to the proof of the \textit{statement 2} and our reformulation \eqref{reformulationRT} of the RT formula.
	{We emphasize that the embeding of $\mathrm{d}\Sigma$ into $\mathcal{E}_{\mathcal{A}}$ is purely out of computational convenience, which allows us to remove the absolute sign in the integration and perform it with ease. In principle, \eqref{density1} is independent of the assignment of the orientations of $V_{\vec x}^\mu$. 
	}

\begin{figure}
		\centering
		\includegraphics[scale=0.40]{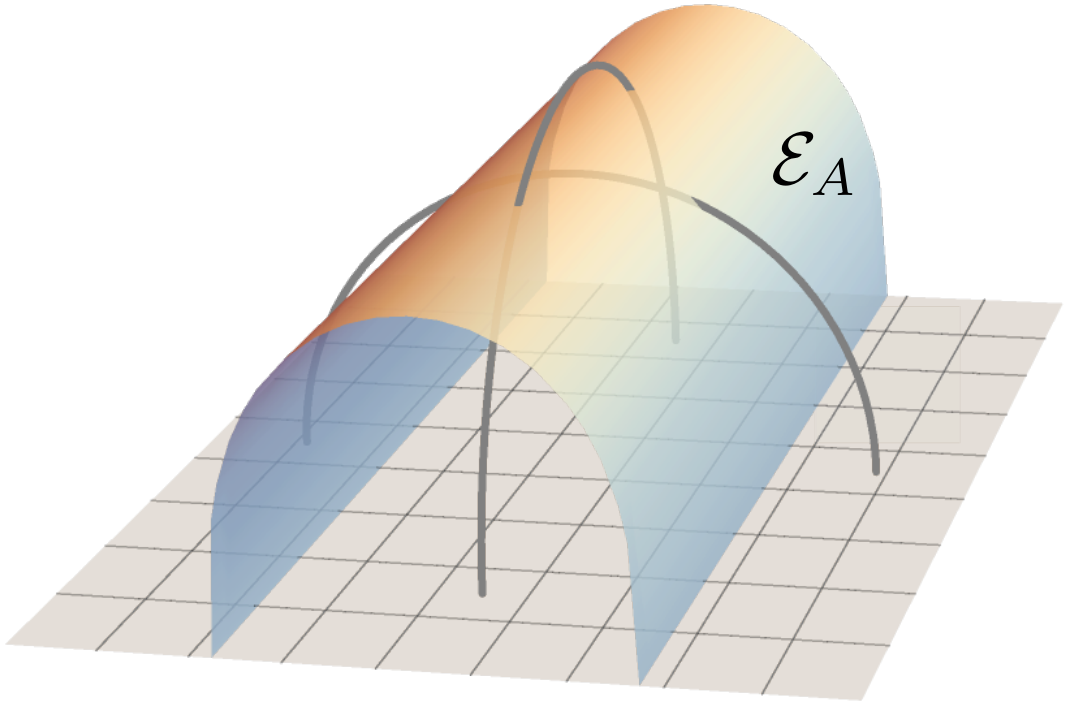}
		\caption{The RT surface $\mathcal{E}_A$ for a strip region $A$ in Poincar\'e AdS$_4$.
			The PEE threads (the gray curves) with both endpoints inside or outside $A$ can possibly pass through $\mathcal{E}_A$ twice.
		}
		\label{fig:stripRT}
	\end{figure}

	One may wonder whether the above derivation still is valid, if we embed $\D\Sigma$ on the RT surface of a non-spherical region $\cal A$ and assign positive direction to these threads from $\mathcal{A}$ to $\bar{\cal A}$, for example in a strip. The answer is no.
	In Fig.\ \ref{fig:stripRT} we show the PEE threads passing through the RT surface for a strip region. 
	We see that there are PEE threads emanating from the strip and passing through $\mathcal{E}_{A}$ twice, one from the inside to the outside and the other from the outside to the inside.
	For any $\D\Sigma$ on the RT surface of the strip, the whole set of PEE threads emanating from the strip with the initial orientation pointing inward the bulk does not satisfy \eqref{propertyB}.
	That is, the set of PEE threads emanating from the strip does not all pass through $d\Sigma$ from one side to the other, hence we cannot remove the absolute value in the integration: 
	\begin{align}
		\textit{density of intersections}=\frac12\int_{\partial \mathcal{M}}\D ^{d-1}\vec x|V_{\vec x}^\mu(Q) n_\mu(Q)|\,~~\neq~~\int_{\rm strip}\D ^{d-1}\vec x V_{\vec x}^\mu(Q) n_\mu(Q)\,,
	\end{align}
	In other words, there are threads passing through $\D\Sigma$ with opposite orientation hence cancel with each other in the above integration, hence this integration does not capture the number of the intersections.

	\section{Reconstruction beyond the RT surfaces}\label{sec4}
	Previously we proved the statement 2, and use it to give a reformation of the RT formula in AdS space. Actually, the \textit{statement 2} is very powerful and enough to generalize our reconstruction from the RT surfaces to the area of a generic co-dimension 2 hyper-surface in the AdS bulk.\footnote{By co-dimension 2 hypersurfaces we mean co-dimension 2 hypersurfaces in AdS spacetime.} In other words, we can interpret the area of all the geometric quantities in terms of the boundary PEEs. The generalization is straightforward, given an arbitrary co-dimension 2 surface $\Sigma$ we have
	\begin{align}\label{countingintersecionsSigma}
		\frac{\mathrm{Area}[\Sigma]}{4G}=&	\mathcal{N}(\Sigma)\notag\\
		=&\frac12
		\int_{\partial \mathcal{M}} \D^{d-1}\vec x\int_{\partial \mathcal{M}}\D^{d-1}\vec y~\omega_{\Sigma}(\vec x,\vec y)\mathcal I(\vec x,\vec y)\,.
	\end{align}
	Here $\Sigma$ does not have to be a surface homologous to any boundary region. Since the density of intersections on any infinitesimal area elements in the bulk is always $1/(4G)$, one can divide $\Sigma$ into infinitesimal area elements $\D\Sigma$, calculate the number of intersections $\text{Area}(\D\Sigma)/(4G)$ and finally sum over all the area elements to get the total number of intersections $\mathcal{N}(\Sigma)$ on $\Sigma$, which is just given by $\text{Area}(\Sigma)/(4G)$.
	
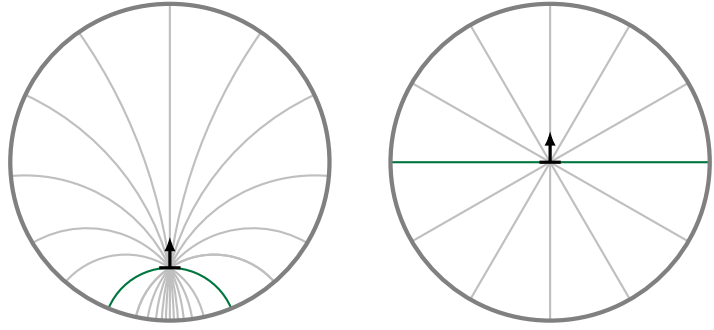
\begin{figure}
		\centering
		\begin{tikzpicture}[scale=0.7]
			\clip(-3.5,-3.2)rectangle(3.5,3.2);
			\draw[thick,LouisColor1](-90-22.6199:3)arc(90+67.3801:90-67.3801:1.25);
			\draw[thick,gray!50](-90-8.03116:3)arc(180-8.03116:{180-2*56.3099-8.03116}:2);
			\draw[thick,gray!50](-90+8.03116:3)arc(8.03116:{+2*56.3099+8.03116}:2);
			\draw[thick,gray!50](0,-3)--(0,3);
			\draw[thick,gray!50](-90+4.99892:3)arc(4.99892:{+2*45.+4.99892}:3);
			\draw[thick,gray!50](-90-4.99892:3)arc(180-4.99892:{180-2*45.-4.99892}:3);
			\draw[thick,gray!50](-90-12.2528:3)arc(180-12.2528:{180-2*63.4349-12.2528}:1.5);
			\draw[thick,gray!50](-90-12.2528:3)arc(180-12.2528:{180-2*63.4349-12.2528}:1.5);
			\draw[thick,gray!50](-90+12.2528:3)arc(12.2528:{2*63.4349+12.2528}:1.5);
			\draw[thick,gray!50](-90+2.91038:3)arc(2.91038:{2*30.9638+2.91038}:5);
			\draw[thick,gray!50](-90-2.91038:3)arc(180-2.91038:{180-2*30.9638-2.91038}:5);
			\draw[thick,gray!50](-90-1.43796:3)arc(180-1.43796:{180-2*16.6992-1.43796}:10);
			\draw[thick,gray!50](-90+1.43796:3)arc(1.43796:{2*16.6992+1.43796}:10);
			\draw[black,very thick](-0.2,-2)--(0.2,-2);
			\draw[black,-{Latex[length=2mm]},very thick](0,-2)--(0,-1.4);
			\draw[ultra thick,gray](0,0)circle(3);
		\end{tikzpicture}
		\begin{tikzpicture}[scale=0.7]
			\clip(-3.5,-3.2)rectangle(3.5,3.2);
			\draw[LouisColor1,thick](3,0)--(-3,0);
			\draw[gray!50,thick](0,3)--(0,-3);
			% \draw[LouisColor1,thick](45:3)--(225:3);
			\draw[gray!50,thick](30:3)--(30+180:3);
			\draw[gray!50,thick](60:3)--(60+180:3);
			\draw[gray!50,thick](-30:3)--(-30+180:3);
			\draw[gray!50,thick](-60:3)--(-60+180:3);
			% \draw[LouisColor1,thick](135:3)--(-45:3);
			\draw[black,very thick](-0.2,0)--(0.2,0);
			\draw[black,-{Latex[length=2mm]},very thick](0,0)--(0,0.6);
			\draw[ultra thick,gray](0,0)circle(3);
		\end{tikzpicture}
		\caption{The area element $\D\Sigma$ and its direction is represented by the black arrow. The gray curves represent all the PEE threads that involve in the reconstruction of $\D\Sigma$, and the green curve represents the PEE thread that saturates the lower bound.}
		\label{fig:globalAdS}
	\end{figure}

	Interestingly, given the position of an infinitesimal area element $\D\Sigma$ at $Q$, all the PEE threads passing through $Q$ contribute to the reconstruction of $\D\Sigma$. 
	It is easy to see that, the scale $|\vec x-\vec y|$ of all the two-point PEEs (or PEE threads) $\mathcal{I}(\vec x,\vec y)$ that contribute to the reconstruction of $\D\Sigma$, is lower bounded. For example, in Poincar\'e AdS$_{d+1}$, the scale of those two-point PEEs satisfies
	\begin{align}
		|\vec x-\vec y|\geq 2 z_0,
	\end{align}
	where $z_0$ is the $z$ coordinate of $\D\Sigma$. In other words, the two-point PEEs with $|\vec x-\vec y|< 2 z_0$ will not contribute to the reconstruction of area elements deeper than $z_0$. 
	See Fig.\ \ref{fig:globalAdS} for two examples in global AdS$_{3}$, where we show the sets of PEE threads that reconstruct two different area elements.
	In the left case, the $\D\Sigma$ is close to the boundary, hence the small scale PEEs are involved. 
	In the right case, the $\D\Sigma$ is in the center of the AdS space and only the largest scale PEEs are involved in its reconstruction. At the same time given the position of $Q$, each PEE thread that passes through $Q$ will give different contributions to the area of $\D\Sigma$ if we vary the direction of $\D\Sigma$. Since $\D\Sigma$ is infinitesimal, this reconstruction could be regarded as the reconstruction for the bulk point $Q$ with a direction $n^{\mu}(Q)$.

	\section{Equivalence to the Crofton formula}\label{sec5}
	The main result \eqref{countingintersecionsSigma} in this paper indeed indicates a pure mathematical theorem if we cancel the $1/(4G)$ coefficient\footnote{Note that there is a factor $c/6=1/(4G)$ in $\mathcal{I}(\vec x, \vec y)$.} on both sides. 
	It claims that, when the density of geodesics in AdS space is properly defined, one can calculate the area of any co-dimension 2 surface by counting the number of intersections between the surface and all the geodesics in AdS. 
	Remarkably, similar mathematical statements for a generic Riemannian manifold have been developed in integral geometry, which is called the \emph{Crofton formula} 
	\footnote{{See \ref{sec:crofton} for more details of the Crofton formula. A comprehensive review of the classical integral geometry is provided in \cite{santalo2004integral}}}. 
	
	{
	The Crofton formula states that given a $d$-dimensional Riemannian manifold $M$ and a $(d-1)$-dimensional hypersurface $\Sigma$, then the area of $\Sigma$ is given by
	\begin{equation}
		\frac12 \frac{d-1}{\Omega_{d-2}}\int_{\text{Geod}(M)}\#(\Sigma\cap \Gamma)\D  \Gamma = \mathrm{Area}(\Sigma),
	\end{equation}
	where $\Omega _{d-1}= \frac{2\pi^{{d-1}/2} }{\Gamma((d-1)/2)}$ is the area if the $(d-2)$-dimensional sphere, $\#(\Sigma \cap \Gamma)$ is the number of intersections of $\Sigma$ and a geodesic $\Gamma$ and $\D \Gamma$ is a properly chosen measure over the space of geodesics $\text{Geod}(M)$. The measure $\D \Gamma$ has to be invariant under changes of local coordinates and invariant under translations along geodesics. More explicitly, given the metric $g_{\mu\nu}$ of the Riemannian manifold $M$, we put
	\begin{equation}
		\mathcal{L} = \sqrt{g_{\mu\nu}\dot{x}^\mu \dot{x}^\nu},\quad p_\mu = \frac{\partial \mathcal{L}}{\partial \dot{x}^\mu},
	\end{equation} 
	and $\dot{x}^\mu := \D x^\mu(\tau) / \D\tau$, with some parameters $\tau$ of geodesics. Then the measure $d\Gamma$ is
	\begin{equation}
	\D \Gamma = (\D p_\mu \wedge\mathrm{d}x^\mu)^{d-1}.
	\end{equation}
	}
	
	Now we will show our result \eqref{countingintersecionsSigma} is equivalent to the Crofton formula.

	Consider the length of the geodesic in AdS space between two boundary points $\vec x_1$ and $\vec x_2$,
	\begin{equation}
	\ell(\vec x_1,\vec x_2) = \min_{\vec x_1 \rightarrow  \vec x_2}\int^{\tau_2}_{\tau_1} \D \tau \ \mathcal{L} = \min_{\vec x_1 \rightarrow  \vec x_2}\int^{\tau_2}_{\tau_1}  \D \tau \sqrt{g_{\mu\nu} \dot{\gamma}^\mu \dot{\gamma}^\nu},
	\end{equation}
	where $\gamma ^\mu(\tau)$ is a path from $\vec x_1$ to $\vec x_2$. 
	With the standard Hamilton-Jacobi argument, the conjugate momentum is given by $p_\mu |_{\vec x_2}= \partial \ell(\vec x_1,\vec x_2) / \partial x_2 ^\mu$.
	
	The key feature of pure AdS is that all geodesics have two endpoints on the boundary, and there is one and only one geodesic connecting a given pair of boundary points.
	Therefore, we are free to parameterize the geodesics using its endpoints on the boundary. 
	Now we restrict $x_1$ and $x_2$ to the boundary, and parameterize the geodesics using its endpoints $(\vec{x}_1,\vec{x}_2)$, with coordinates $x_{1,2}^i,i=1, \ldots ,d-1$. 
	The invariant density of geodesics is given by the $(2d-2)$-form \eqref{eq:inv_dens}.
	In terms of boundary coordinates $(\vec x_1,\vec x_2)$, the density becomes
	\begin{equation}\label{eq:Gamma_ads}
	\D \Gamma = \det \left( \frac{\partial^2 \ell (\vec{x}_1,\vec{x}_2)}{\partial \vec{x}_1 \partial \vec{x}_2} \right) \D  \vec{x}_1 \wedge \D \vec{x}_2,
	\end{equation}
	where $\D \vec x_{1,2}=\D x^1_{1,2}\wedge\D x^2_{1,2}\wedge\cdots\wedge \D x^{d-1}_{1,2}$. 
	With $\ell(\vec{x}_1,\vec{x}_2)=2 \log  \frac{|\vec{x}_1 - \vec{x}_2|}{\epsilon} $ in pure AdS, we have
	\begin{equation}
	\det \left( \frac{\partial^2 \ell (\vec{x}_1,\vec{x}_2)}{\partial \vec{x}_1 \partial \vec{x}_2} \right) = \det\left( 2\frac{-\delta ^{ij}|\vec{x}_{1} -\vec{x}_{2}|^{2} +2\left( x_{1}^{i} -x_{2}^{i}\right)\left( x_{1}^{j} -x_{2}^{j}\right)}{|\vec{x}_{1} -\vec{x}_{2}|^{4}}\right) = \frac{2^{d-1}}{|\vec{x}_1 - \vec{x}_2|^{2d -2}}.
	\end{equation}
	Since $c=3 /2 G$, and
	\begin{equation}
	\mathcal{I}(\vec{x}_1,\vec{x}_2)=\frac c6\frac{2^{d-1}(d-1)}{\Omega_{d-2}|\vec{x}_1-\vec{x}_2|^{2(d-1)}}=\frac{c (d-1)}{6\Omega_{d-2}}	\det \left( \frac{\partial^2 \ell (\vec{x}_1,\vec{x}_2)}{\partial \vec{x}_1 \partial \vec{x}_2} \right),
	\end{equation}
	it is easy to find that the Crofton's formula \eqref{eq:Crofton} can be written as
	\begin{equation}\label{eq:pee-crofton}
	\frac{\mathrm{Area}(\Sigma)}{4G}=\frac12\int_{\partial\mathcal{M}}\mathrm{d}^{d-1}\vec{x}_1\int_{\partial\mathcal{M}}\mathrm{d}^{d-1}\vec{x}_2\mathrm{~}\omega_\Sigma(\vec{x}_1,\vec{x}_2)\mathcal{I}(\vec{x}_1,\vec{x}_2),
	\end{equation}
	which is exactly \eqref{countingintersecionsSigma}. Here $\omega_\Sigma(\vec{x}_1,\vec{x}_2)$ is just $\#(\Sigma\cap \Gamma)$ with $\Gamma$ being the PEE thread connecting the pair of boundary points $\{\vec x_1,\vec x_2\}$. Hence our main result \eqref{countingintersecionsSigma} is equivalent to the \emph{Croftron formula} in AdS space. 
	
	In addition, beyond pure AdS space, if any geodesic can be parameterized by boundary two points, then the area of $\Sigma$ is expressed by
	\begin{equation}
	\frac{\mathrm{Area}(\Sigma)}{4G}=\frac12 \frac{d-1}{\Omega_{d-2}} \int_{\partial\mathcal M}\D^{d-1}\vec x_1\int_{\partial\mathcal M}\D^{d-1}\vec x_2 \omega_{\Sigma}(\vec x_1,\vec x_2)\det \left(
	\frac{\partial^2\ell(\vec x_1,\vec x_2)}{\partial\vec x_1\partial\vec x_2}
	\right).
	\end{equation}

\section{Discussions}\label{sec6}
\subsection{Conclusion}
In summary, based on the PEE structure of a CFT and its geometrization scheme using the bulk geodesics anchored on the boundary, we can obtain a network of geodesics (or PEE threads) with the density of geodesics specified everywhere in the bulk \cite{Lin:2023rbd}. 
In this paper we showed that, if we cut the AdS space open along any co-dimension 2 surface $\Sigma$, the density of cuts of the bulk geodesics is always $1/4G$ at any point on $\Sigma$. 
In other words, the AdS is woven by geodesics, with the geodesics distributed uniformly at any point and along any direction. 
Then the area of any co-dimension 2 surface in AdS can be reconstructed by counting the number of intersections between the surface and the PEE network.
	
Based on the PEE network, we provide a reformulation of the RT formula, which identifies the RT surface of an boundary region $A$ as the surface $\Sigma_{A}$ homologous to $A$ and has the minimal number of intersections with the PEE network. 
Our reformulation of the RT formula is partially inspired by the computation of the entanglement entropy in tensor network models, which is given by minimal number of cuts between a homologous path and the tensor network. Here the PEE network is analogous to the tensor network, and the number of intersections $\mathcal{N}(\Sigma_A)$ is akin to the number of cuts in the tensor network. The PEE network can be viewed as a tensor network that precisely captures the entanglement structure of the boundary CFT at large $c$ limit.
	
Remarkably, our reconstruction of surface areas in AdS space has a pure mathematical core, namely the \textit{Crofton formula} from integral geometry. 
This can be proved by showing that the PEE structure $\mathcal{I}(\vec x,\vec y)\D\vec x \D\vec y$ coincides with the \emph{kinematic measure}, the invariant measure on the space of geodesics in AdS space. 
Despite the formal coincidence, our proposal is physically motivated: to understand the bulk geometry in terms of the boundary entanglement structure. 
Moreover, our derivation is totally different from the \textit{Crofton formula} in mathematics. 
In our proposal, every bulk geodesic has a physical meaning, representing a two point PEE in the boundary CFT. For any given bulk surface $\Sigma$, we can identify the class of two-point PEEs in the dual CFT that gives non-trivial contribution to the area of $\Sigma$ by tracing the class of PEE threads passing through $\Sigma$. 
Furthermore, the magnitude of contribution from each PEE thread is proportional to the corresponding two-point PEE $\mathcal{I}(\vec x,\vec y)$, as well as the number $\omega(\vec x,\vec y)$ and direction of the intersection with $\Sigma$.

\subsection{Relation to other research drections}
\textbf{The extensive mutual information: }

It is worth mentioning that, in \cite{Casini:2008wt} the authors introduced the so-called extensive mutual information (EMI), which is indeed the PEE we have defined if we do not require this quantity to be a mutual information \cite{Han:2019scu}. 
The EMI is also defined to satisfy the set of physical requirements we used to define the PEE. 
In addition, the authors of \cite{Casini:2008wt} assumed that the EMI that solves these requirements should be a mutual information, which is the essential difference between the EMI and the PEE. Nevertheless, the PEE obtained by solving the physical requirements only coincide with the mutual information in two-dimensional free fermions.
In more generic theories, the PEE $\mathcal{I}(A,B)$ is not the mutual information, hence the EMI does not exist \cite{Agon:2021zvp,Wen:2019iyq,Han:2019scu}.
	
Together with \cite{Lin:2023rbd}, we proposed two crucial new understandings of the PEE. 
Firstly, the naive normalization property \eqref{eq:norm} of the PEE only applies to spherical regions, and is wrong for more generic regions. 
Secondly, the two-point PEEs that contribute to $S_{A}$ are not confined in the PEE between points in $A$ and $\bar A$ respectively. 
Actually the contribution is determined by the number of intersections between the PEE threads and the RT surface $\mathcal{E}_{A}$. 
These new understandings provide a solution to an puzzle on EMI proposed by Casini and Huerta in \cite{Casini:2008wt} (see also \cite{Han:2019scu}).   
In \cite{Casini:2008wt}, the authors used the normalization property \eqref{eq:norm} of EMI to evaluate the entanglement entropy for an annulus in the vacuum of holographic CFT$_3$ that is dual to Poincar\'e AdS$_4$. 
The RT formula tells us that the RT surface undergoes a phase transition between two phases with connected or disconnected RT surfaces respectively. 
However, if we naively apply \eqref{eq:norm} to compute the entanglement entropy, there is no such phase transition, and the result of the RT formula can not be reproduced. 
As a result, the authors of \cite{Casini:2011kv} concluded that the EMI does not exist in holographic CFTs. 
In contrast, our scheme reproduces the RT formula for annulus perfectly, by abandoning the requirement that the PEE is a mutual information, modifying the normalization property \eqref{eq:norm} and taking into account the weight of the PEE threads.
More specifically, in our scheme PEE threads across the annulus (i.e. PEE threads stretching between the region surrounded by the annulus and the region outside the annulus) could contribute doubly to the entanglement entropy of the annulus in the disconnected phase, which is absolutely inconsistent with the naive normalization property \eqref{eq:norm}.

\textbf{Differential entropy and Kinematic space}

Although the computations and analyses in this paper may appear simple, we stress that the main results, i.e. the \textit{statement 2} and the reformulation of the RT formula for generic boundary regions, are a significant progress compared with our previous paper \cite{Lin:2023rbd}. 
On the other hand, our analysis gives another derivation of the \textit{Crofton formula} in AdS from a purely physical motivation. 
It is also worth mentioning that, there have already been extensive studies (see \cite{Czech:2015qta,Huang:2015xca,Zhang:2016evx,Czech:2017zfq,Huang:2019ajv,Huang:2020zxh,Huang:2019wzc,Huang:2020cye,Huang:2021qkm} for examples) on exploring holographic geometry reconstruction via the \textit{Crofton formula}. 
Especially in \cite{Czech:2015qta} it was shown that, in AdS$_3$ the differential entropy formula can be derived from the Crofton formula, and any bulk geodesic indeed represents the conditional mutual information between the degrees of freedom on the two boundary endpoints \footnote{Since it has been shown in \cite{Rolph:2021nan} that, the conditional mutual information in CFT$_2$ is exactly the PEE computed by the ALC proposal, our results exactly reproduce those in \cite{Czech:2015qta}. }. Nevertheless, most of these analyses in the above literature are confined to a time slice in AdS$_3$. 
It was unclear how to represent these geodesic using entropy quantities in higher dimensions, and here we give a satisfactory answer to this question: all the geodesics in the \textit{Crofton formula} are represented by the two-point PEEs on the boundary CFT. Our work (at least) gives another formalism, or perhaps more natural formalism by generalizing the story to higher dimensions in a natural way, for the area reconstruction in holography based on the partial entanglement entropy.
	
Although our discussion is confined to a static time slice of Poincar\'e AdS, it can reconstruct the area of the RT surfaces for a generic boundary region in general dimensions.  
Our configuration can also be used to reconstruct the area of a generic co-dimension 2 surface, which goes beyond the RT formula. Furthermore, we can extract more information from the PEE thread configurations than the entanglement entropy, for example, the mixed state correlation represented by the EWCS \cite{Wen:2021qgx} and the entanglement contour\cite{Chen:2014}. From the above discussions, we highlight three fundamental distinctions between our framework and kinematic space approach \cite{Balasubramanian:2013lsa,Headrick:2014eia,Czech:2014wka,Czech:2014ppa,Czech:2015qta,Czech:2015kbp}:
\begin{itemize}
\item \textit{Conceptual foundations:}
The studies on Kinematic space are based on the Crofton formula in integral geometry, whereas our approach in some sense re-derived the Crofton formula in Poincare AdS. The two-point PEEs on the boundary are natural physical interpretation for the geodesics in Crofton formula.
\item \textit{Network interpretations:}
In the previous papers on Kinematic space, the discussion is mainly focused in the Kinematic space of the geodesics, while our discussion is mainly focused in the AdS space. In \cite{Czech:2015kbp}, the MERA tensor network was argued as a discretized embedding in the kinematic space. While in our construction, we take the PEE network as a continuous tensor network that make up the Poincare AdS space.
\item \textit{Higher dimension generalizations:}
In the previous papers on Kinematic space, while a geodesic in Poincare AdS$_3$ space could be interpreted as a two-point conditional mutual information (CMI), the higher-dimensional generalizations are missing. Our framework resolves this by establishing a correspondence: the geodesics in general dimensional AdS space could be interpreted as the two-point PEEs, which coincide with two-point CMI in Poincare AdS$_3$.
\end{itemize}

\subsection{Future directions}
In a later paper \cite{Wen:2024uwr}, the PEE thread configurations has been generalized to holographic state in island phase. 
In the following we give an incomplete list of future directions.
	
\textbf{Going beyond AdS space:} Although our discussion is confined to pure AdS spaces, the \textit{Crofton formula} applies to generic Riemannian manifolds. For a generic Riemannian manifold and the corresponding \emph{kinematic measure}, it is natural to take the geodesics as a representation of the two-point PEE in the dual boundary field theory, hence interpreting the area of the bulk co-dimension one surfaces in terms of the boundary entanglement structure. On the other way around, the bulk PEE thread configuration is also useful to determine the PEE structure for states where the PEE can not be solved from the physical requirements.

{It is interesting to observe that, the PEE threads emanating from the boundary fully foliate the Poincar\'e AdS space. But we think this is not an coincidence. The underlying reason for this complete foliation may be that the boundary state is pure, allowing the bulk geometry to be fully reconstructed from boundary data. Moreover, all threads must return to the boundary, as there should be no open endpoints in the bulk. If we extend the analysis to a BTZ black hole, threads originating from the boundary can only fully foliate the region outside the horizon (see forthcoming work \cite{Wen:2026}). In this case, some threads will enter the horizon, terminating at the singularity or extending to the second boundary in an eternal black hole setup. This reflects the thermal nature of the boundary state, with threads reaching the horizon contributing to the thermal entropy.}

\textbf{Covariant configurations:} An important step is to go beyond static space (or Riemannian manifold), and reconstruct the area of surfaces in spacetime. This seems achievable since there is a successful covariant of the RT formula \cite{Hubeny:2007xt}, which is also proved \cite{Dong:2016hjy}. For this purpose, we may need to study the covariant version of the PEE structure and using the HRT surface to represent the two-point PEEs. 
	
\textbf{Reconstructing general dimensional submanifolds:} 
In the main text, we only discussed the reconstructing the area of co-dimension one bulk surfaces.
Actually the reconstruction can be generalized to lower dimensional submanifolds. 
The \textit{Crofton formula} also applies to codimension-$n$ ($n\geq 2$) submanifolds \cite{paiva2008gelfand}:
	\begin{equation}
		\text{Vol}(N) = c_{n,d} \int _{\text{Geod}_{n}(M)} \#(N \cap  \Gamma) \mathrm{d} \Gamma ,
	\end{equation} 
	where $\text{Geod}_n(M)$ is the space of all $n$-dimensional totally geodesic submanifolds $\Gamma$ and $c_{n,d}$ is some constant depending on the dimension of manifold $M$ and the dimension of the codimension-$n$ object $N$. 
    The existence of totally-geodesic submanifolds of dimension $n\geqslant 2$ in a general Riemannian manifold is exceptional. 
    But with the help of symmetries, we can find the measure $\mathrm{d} \Gamma$ and calculate the volume of $N$ with the Crofton formula, see \cite{Czech:2019hdd} for an example in AdS$_3$.

	\textbf{PEE network as a tensor network toy model of quantum gravity}: Compared to previous tensor network toy models of gravity, the PEE network is a well-defined continuous network that naturally extends to higher dimensions. 
    It will be interesting to add bulk degrees of freedom to the PEE network to study quantum corrections \cite{Faulkner:2013ana} and quantum error correction properties \cite{Almheiri:2014lwa,Pastawski:2015qua,Harlow:2016vwg} of the PEE network. {We refer to \cite{Wen:2025gui} for a recent development in this direction. The authors propose three new PEE tensor network models: namely, a factorized PEE, a Happy-like, and a random PEE tensor network. Furthermore, they prove that the number of intersection $\mathcal{N}_{\mathcal{E}_A}$ not only reproduce the area of the RT surface, but also yields the entanglement entropy for any boundary region.
  }

{
\textbf{Bulk metric reconstruction}:
In our framework, the bulk metric is given \emph{a prior}, so that we can define a measure for PEE threads.
However, one of the important problems in AdS/CFT is to derive bulk metric purely from the boundary CFT data.
See \cite{Czech:2014ppa,Bao:2019bib,Bao:2024hwy,Jiang:2024hjz,Jiang:2024xqz} for relevant studies.
It is possible that, matter fields or singularities can act as sources or sinks for PEE threads. Conversely, introducing such sources while still requiring a complete foliation could constrain the background metric in a manner consistent with Einstein’s equations, hence shed light on the bulk metric reconstruction program. We leave this for future investigations.
}

	\section*{Acknowledgements}
		Y. Lu is supported by Shanghai Institute for Mathematics and Interdisciplinary Sciences.
		J. Lin is supported by the National Natural Science Foundation of China under Grant No.12505063. Q. Wen is supported by NSFC Grant No. 12447108.
		Q. Wen would like to thank Bartlomiej Czech, Veronika E. Hubeny, Ling-Yan Hung, Huajia Wang and Zhenbin Yang for helpful discussions.

	% \begin{figure}
		%     \centering
		%     \includegraphics[width=0.4\paperwidth]{Demo.png}
		%     \caption{The locally tangential hemisphere (green) of $\Sigma$ (white curve) at $\lambda$. }
		%     \label{fig:tan_hemi}
		% \end{figure}

	% \begin{figure*}[t]
		% \centering
		% \input{TNW}
		% \caption{The tensor network as the discretized version of PEE threads description in AdS$_3$/CFT$_2$.
			% }
		% \label{fig:TNW}
		% \end{figure*}

	% \begin{acknowledgments}

\begin{appendices}
\section{The Crofton formula}\label{sec:crofton}

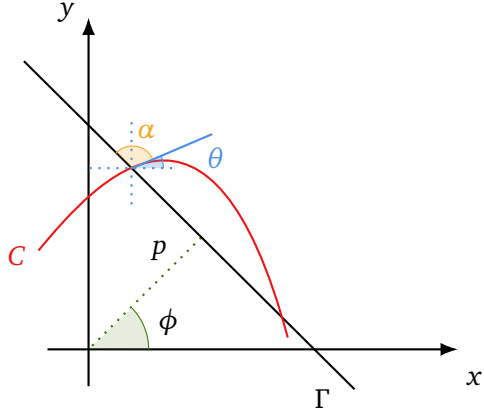
\begin{figure}[t]
\centering
\begin{tikzpicture}[x=0.75pt,y=0.75pt,yscale=-1,xscale=1]
    
    %Shape: Axis 2D [id:dp7735272016788031] 
    % \draw [thick] (41.75,194.95) -- (246.75,194.95)(62.25,28) -- (62.25,213.5) (239.75,189.95) -- (246.75,194.95) -- (239.75,199.95) (57.25,35) -- (62.25,28) -- (67.25,35)  ;
    \draw [thick,-{Latex}] (41.75,194.95) -- (246.75,194.95)  ;
    \draw [thick,-{Latex}](62.25,213.5) --(62.25,28);
    %Straight Lines [id:da05973838844794788] 
    \draw [thick][color={rgb, 255:red, 65; green, 117; blue, 5 }  ,draw opacity=1 ][fill={rgb, 255:red, 126; green, 211; blue, 33 }  ,fill opacity=1 ] [dash pattern={on 0.84pt off 2.51pt}]  (118.5,138.7) -- (62.25,194.95) ;
    %Straight Lines [id:da9101496869532395] 
    \draw  [thick]  (30,50) -- (195,215) ;
    %Shape: Arc [id:dp16019703005839814] 
    \draw [thick] [draw opacity=0][fill={rgb, 255:red, 65; green, 117; blue, 5 }  ,fill opacity=0.13 ] (83.98,174.27) .. controls (89.11,179.65) and (92.25,186.93) .. (92.25,194.95) .. controls (92.25,195.18) and (92.25,195.4) .. (92.24,195.63) -- (62.25,194.95) -- cycle ; \draw  [color={rgb, 255:red, 65; green, 117; blue, 5 }  ,draw opacity=1 ] (83.98,174.27) .. controls (89.11,179.65) and (92.25,186.93) .. (92.25,194.95) .. controls (92.25,195.18) and (92.25,195.4) .. (92.24,195.63) ;  
    %Curve Lines [id:da8171596826737939] 
    \draw [thick][color={rgb, 255:red, 240; green, 22; blue, 22 }  ,draw opacity=1 ]   (37.25,145.25) .. controls (66.75,109.25) and (123.67,46.33) .. (161.67,189) ;
    %Straight Lines [id:da45937211906737474] 
    \draw [thick][color={rgb, 255:red, 74; green, 144; blue, 226 }  ,draw opacity=1 ] [dash pattern={on 0.84pt off 2.51pt}]  (62.7,103.8) -- (105.5,103.8) ;
    %Straight Lines [id:da3897062642047173] 
    \draw [thick][color={rgb, 255:red, 74; green, 144; blue, 226 }  ,draw opacity=1 ] [dash pattern={on 0.84pt off 2.51pt}]  (83.7,122) -- (83.7,80.3) ;
    %Straight Lines [id:da8889195653298356] 
    \draw [thick][color={rgb, 255:red, 74; green, 144; blue, 226 }  ,draw opacity=1 ]   (84.1,103.8) -- (124,86.67) ;
    %Shape: Arc [id:dp03054768197070934] 
    \draw [thick] [draw opacity=0][fill={rgb, 255:red, 74; green, 144; blue, 226 }  ,fill opacity=0.27 ] (98.39,97.42) .. controls (99.22,99.45) and (99.59,101.63) .. (99.5,103.8) -- (83.51,103.98) -- cycle ; \draw  [color={rgb, 255:red, 74; green, 144; blue, 226 }  ,draw opacity=1 ] (98.39,97.42) .. controls (99.22,99.45) and (99.59,101.63) .. (99.5,103.8) ;  
    %Shape: Arc [id:dp08315713821096105] 
    \draw [thick] [draw opacity=0][fill={rgb, 255:red, 245; green, 166; blue, 35 }  ,fill opacity=0.21 ] (75.85,96.3) .. controls (77.89,94.06) and (80.83,92.65) .. (84.1,92.65) .. controls (88.61,92.65) and (92.5,95.33) .. (94.25,99.19) -- (84.1,103.8) -- cycle ; \draw  [color={rgb, 255:red, 245; green, 166; blue, 35 }  ,draw opacity=1 ] (75.85,96.3) .. controls (77.89,94.06) and (80.83,92.65) .. (84.1,92.65) .. controls (88.61,92.65) and (92.5,95.33) .. (94.25,99.19) ;  
    
    % Text Node
    \draw (95.64,173.16) node [anchor=north west][inner sep=0.75pt]    {$\phi $};
    % Text Node
    \draw (92.36,138.98) node [anchor=north west][inner sep=0.75pt]    {$p$};
    % Text Node
    \draw (173.67,213.63) node [anchor=north west][inner sep=0.75pt]    {$\Gamma $};
    % Text Node
    \draw (45.33,19.13) node [anchor=north west][inner sep=0.75pt]    {$y$};
    % Text Node
    \draw (249.33,204.8) node [anchor=north west][inner sep=0.75pt]    {$x$};
    % Text Node
    \draw (20.33,141.63) node [anchor=north west][inner sep=0.75pt]  [color={rgb, 255:red, 240; green, 22; blue, 22 }  ,opacity=1 ]  {$C$};
    % Text Node
    \draw (119.9,92.07) node [anchor=north west][inner sep=0.75pt]  [color={rgb, 255:red, 74; green, 144; blue, 226 }  ,opacity=1 ]  {$\theta $};
    % Text Node
    \draw (85.17,80.6) node [anchor=north west][inner sep=0.75pt]  [color={rgb, 255:red, 245; green, 166; blue, 35 }  ,opacity=1 ]  {$\alpha $};

\end{tikzpicture}
\caption{The parameterization of straight lines in $\mathbb{R}^2$.
    The green line is perpendicular to the straight line $\Gamma$, and its length and angular coordinate are given by $p$ and $\phi$ respectively.}
\label{fig:2d_Crofton}
\end{figure}

The original \textit{Crofton formula} was first proved by M. Crofton, which relates the length of a rectifiable curve $C$ in $\mathbb{R}^{2}$ to its intersection number with straight lines $\Gamma$ (geodesics) as
\begin{equation}
\text{Length}(C) = \frac{1}{2} \int _{\overline{\text{Gr}}_1(\mathbb{R}^{2})} \#(C \cap  \Gamma) \mathrm{d} \Gamma,
\end{equation}
where $\overline{\text{Gr}}_1 (\mathbb{R}^{2})$ is the space of geodesics in $\mathbb{R}^{2}$, $\mathrm{d} \Gamma$ is an appropriately normalized measure invariant under rigid motions {(translations and rotations)}, and $\#(C \cap  \Gamma)$ is the number of times a geodesic $\Gamma$ intersects with the curve $C$. The geodesics in $\mathbb{R}^2$ can be parameterized by 
\begin{equation} \label{eq:straight_line}
x\cos\phi+y\sin\phi-p=0,
\end{equation}
where $p$ is the distance from origin and $\phi$ is the direction angle; see Fig.~\ref{fig:2d_Crofton}. The invariant measure on $\overline{\text{Gr}}_1(\mathbb{R}^{2})$ is 
\begin{equation}
\mathrm{d} \Gamma = \mathrm{d} p \wedge \mathrm{d} \phi.
\end{equation}
It is easy to check this measure is invariant under rigid motions.

Assuming that $C$ is parameterized by $x=x(s),y=y(s)$, where the parameter $s$ is the length parameter. Let $\alpha$ denote the angle between the tangent at the point on $C$ and the straight line $\Gamma$ intersect $C$ at this point and $\theta$ the angle between the tangent to $C$ and $x$-axis. The measure expressed with parameters $s$ and $\alpha$ is then
\begin{equation}
    \mathrm{d} \Gamma = | \sin \alpha | \mathrm{d} s \wedge \mathrm{d} \alpha,
\end{equation}
The proof of the Crofton formula is now obvious
\begin{equation}
    \frac{1}{2}\int ^{\text{Length}(C)}_{0} \mathrm{d} s \int ^{\pi}_{0} |\sin\alpha| \mathrm{d}\alpha = \text{Length}(C).
\end{equation}

The Crofton formula can be generalized to Riemannian manifolds. Considering a $d$-dimensional Riemannian manifold $M$ (with a metric $g_{\mu\nu}$) and an arbitrary $(d-1)$-dimensional immersed hypersurface $\Sigma \subset M$, the Crofton formula states
\begin{equation} \label{eq:crofton_Riemann}
\text{Vol}(\Sigma) = c \int _{\text{Geod}(M)} \#(\Gamma \cap  \Sigma) \mathrm{d} \Gamma
\end{equation}
where $\text{Geod}(M)$ is the space of all geodesics in $M$, $c$ is a constant and $\mathrm{d} \Gamma$ is a properly chosen measure on $\text{Geod}(M)$, which has to be invariant under changes of local coordinates and invariant under translations along geodesics.

The derivation of the \textit{Crofton formula} in general Riemannian manifolds follows the similar strategy to that in the $\mathbb{R}^{2}$ case. We decompose the measure $\D \Gamma$ into two parts, one of which represents the position of all the points on the hypersurface $\Sigma$ and the other represents all the tangent direction of the geodesics passing through any point on $\Sigma$.

More explicitly, given an arbitrary curve $x^\mu(\tau)$ in $M$, an invariant scalar (the action) can be constructed by
\begin{equation}
\mathcal{L} = \sqrt{g_{\mu \nu} \dot{x}^\mu \dot{x}^\nu},
\end{equation}
where $\dot{x}^\mu\equiv \D x^\mu(\tau)/\D \tau$.
By definition, the geodesics of $M$ are integral curves of the Euler equation
\begin{equation}\label{eq:geo}
\frac{\D }{\D \tau}\left(
\frac{\partial\mathcal L}{\partial \dot x^\mu}
\right)-\frac{\partial \mathcal L}{\partial x^\mu}=0.
\end{equation}
The conjugate momentum related to $x^\mu$ is $p_\mu \equiv \partial \mathcal{L} / \partial \dot{x}^\mu$ and is subject to the constraint $p^2  =1$, i.e. it is a unit covariant vector in the cotangent space of $x^\mu(\tau)$.

The co-sphere bundle $S^* M$ (i.e. pair $(x^\mu, p_\mu)$ with $p^2 =1$) forms a manifold  with $\dim =2d-1$, which is the phase space of $ M$.
And $(x^{\mu},p_\mu)$ serves as a local coordinate system for $S^* M$, called the canonical coordinates.
The geodesics $\Gamma$ of ${M}$ define a foliation $F_\Gamma$ of $S^* M$, in which the leaves $S^* M/F_\Gamma$ with dimension $2d-2$. 
The canonical two-form
\begin{equation}
\D p_\mu\wedge \D x^\mu
\end{equation}
on $S^* M$ is invariant under local coordinate transformations and the displacements on the leaves of the foliation (i.e. independent of the parameter $\tau$) \cite{santalo2004integral}.
It follows that the following $(2d-2)$-form on $S^* M$ is also invariant
\begin{equation}
\begin{aligned}
    \D \Gamma &= (\D p_\mu \wedge \D x^\mu)^{d-1} \\ &= \sum ^{d}_{i=1} \D p_1 \wedge \D x^1 \wedge\cdots \wedge \D p_{i-1} \wedge \D x^{i-1} \wedge \D p_{i+1} \wedge \D x^{i+1} \wedge \cdots \wedge \D p_{d}\wedge \D x^{d},
\end{aligned} \label{eq:inv_dens}
\end{equation}
which defines an invariant measure of $(2d-2)$-dimensional set of geodesics $\Gamma$.
It is called \emph{kinematic measure}. Also the space of all the geodesics in the manifold is called the \emph{kinematic space} \cite{Czech:2014ppa,Czech:2015qta,Czech:2015kbp,Czech:2016xec}.

Now we consider an arbitrary co-dimension one surface $\Sigma$ in $M$ and choose the coordinates such that $\Sigma$ is given by $x^d =0$, and $(x_1,x_2\cdots,x_{d-1})$ are the orthogonal coordinates on $\Sigma$. In this case the invariant \emph{kinematic measure} $\D \Gamma$ becomes
\begin{equation}\label{eq:gamma2}
\D \Gamma = \D p_1 \wedge \D x^1 \wedge\cdots \wedge \D p_{d-1}\wedge \D x^{d-1}.
\end{equation}
In other words, the bulk geodesics are parameterized by the position where they intersect with $\Sigma$ and the intersection direction captured by the $d-1$ independent conjugate momentum. 
To perform the integral over $\D \Gamma$, note that the conjugate momentum can be parameterized by the angle $\alpha _i$ between the momentum and $x^i$ axis: $p_i = \sqrt{g_{i i}} \cos \alpha _i$. Up to the sign, the \emph{kinematic measure} becomes
\begin{equation}
\D \Gamma = \left( \prod^{d-1}_{i=1}\sqrt{g_{i i}} \sin\alpha _i \right) \D x^1 \wedge \cdots \wedge \D x^{d-1}\wedge \D \alpha _{1} \wedge \cdots \wedge \D \alpha _{d-1},
\end{equation}
where we recognize the area element $\D \sigma$ of a $(d-1)$-dimensional hypersurface $\Sigma$:
\begin{equation}
\D \sigma = \left(
\prod_{i=1}^{d-1} \sqrt{g_{i i}}
\right) \D x^1\wedge \cdots \wedge \D x^{d-1},
\end{equation}
and the area element on the unit $(d-1)$-dimensional sphere in the direction $p_d$
\begin{equation}
\D \Omega_{d-1} = \frac{\sin \alpha _1 \cdots  \sin \alpha _{d-1}}{\cos \alpha _d} \D \alpha _1\wedge \cdots \wedge\D \alpha _{d-1}.
\end{equation}
Hence we can write
\begin{equation}
\D \Gamma = |\cos \alpha _d|  \D \sigma \wedge \D \Omega_{d-1},
\end{equation}
and perform integral for the form $\D \Gamma$ over $\Sigma$ and the upper half of the $(d-1)$-dimensional unit sphere\footnote{
The integral is over half the unit sphere because the geodesics are unoriented.
}
\begin{equation}
\begin{aligned}\label{eq:crofton1}
    \int_{\Sigma\times \Omega_{d-1}/2}\D \Gamma&=\int _{\Sigma} \D \sigma \int_{\Omega_{d-1}/2} |\cos \alpha_{d} | \D \Omega _{d-1}\\&= \mathrm{Area}(\Sigma) \frac{\Omega_{d-2}}{d-1},
\end{aligned}
\end{equation}
where $\Omega_{d-2}=\frac{2\pi^{(d-1)}/2}{\Gamma((d-1)/2)}$ is the area of the $(d-2)$-dimensional sphere and in the last step we apply the following relation
\begin{equation}
\D \Omega_{d-1}= \sin^{d-2}\alpha_{d}\D \alpha_d \D \Omega_{d-2},
\end{equation}
Now note that each geodesic possibly intersects $\Sigma$ multiple times, then the L.H.S of \eqref{eq:crofton1} can be rewritten as
\begin{equation}
\int_{\Sigma\times \Omega_{d-1}/2}\D \Gamma=\frac12\int_{\text{Geod}(M)}\#(\Sigma\cap \Gamma)\D  \Gamma,
\end{equation}
where the factor $\frac12$ accounts for the two orientations of geodesics, as the integral on the R.H.S is over all oriented geodesics.
Now we obtain the Crofton's formula for codimension-1 surface $\Sigma$ and geodesics $\Gamma$
\begin{equation}\label{eq:Crofton}
\begin{aligned}
    \frac12 \frac{d-1}{\Omega_{d-2}}\int_{\text{Geod}(M)}\#(\Sigma\cap \Gamma)\D  \Gamma &= \mathrm{Area}(\Sigma).
\end{aligned}
\end{equation}
\end{appendices}

	\bibliographystyle{JHEP.bst}
\bibliography{ref.bib}

\end{document}